%% file: paper_compiled.tex
\documentclass[review=false]{jfp-epi}
\usepackage{jfp-cup2epi}

\usepackage{graphicx}

\usepackage{style}

\usepackage{planarforest}
\input{planarforest_tikz.tex}

\begin{document}

\journaltitle{JFP}
\cpr{Cambridge University Press}
\doival{10.1017/xxxxx}

\lefttitle{Arboretum.hs}
\righttitle{Journal of Functional Programming}

\totalpg{\pageref{lastpage01}}
\jnlDoiYr{2025}

\title{Arboretum.hs: Symbolic manipulation for algebras of graphs}

\author{Eugen Bronasco}
\orcid{0009-0006-3904-5523}
\affiliation{
    \institution{Mathematical Sciences,\\
    Chalmers and Gothenburg University}
    \city{Gothenburg}
    \country{Sweden}
    \authoremail{bronasco@chalmers.se}
}

\author{Jean-Luc Falcone}
\orcid{0000-0003-0748-6069}
\affiliation{
    \institution{Department of Computer Science,\\
    University of Geneva}
    \city{Geneva}
    \country{Switzerland}
    \authoremail{Jean-Luc.Falcone@unige.ch}
}

\author{Gilles Vilmart}
\orcid{0000-0003-4593-1012}
\affiliation{
    \institution{Section of Mathematics,\\
    University of Geneva}
    \city{Geneva}
    \country{Switzerland}
    \authoremail{Gilles.Vilmart@unige.ch}
}

\begin{abstract}
We design the Arboretum.hs package for symbolic computations with algebras of
trees and more general graphs in Haskell. Thanks to the declarative nature of functional
programming, the package's implementation closely follows mathematical
definitions, making the code intuitive and transparent for users working with
algebraic and combinatorial structures.

To assist with current mathematical research, Arboretum.hs supports
experimentation by facilitating the introduction of new algebraic operations,
as well as providing functionality for rendering trees and forests through
LaTeX integration. 

Compared to recent imperative implementations in languages such as Julia or Python,
Arboretum.hs offers greater flexibility for manipulating and extending
tree-based structures. Its use of Haskell enables safe programming and strong
compile-time guarantees, serving both as a practical computational tool and a
foundation for further research in algebraic combinatorics, beyond the setting
of trees usually considered in the implementation of Butcher series, which are a
fundamental tool for the analysis of numerical integrators.
\end{abstract}

\maketitle

Many areas of mathematics make extensive use of graphs to encode algebraic and combinatorial structures. This is evident in non-associative algebra, where trees encode the different ways of bracketing products, and more generally in operad theory, where they naturally represent the composition of operations, see \cite{chapotonPreLieAlgebrasRooted2001,LodayAO12}. Similar combinatorial structures arise in rough path theory, as developed in~\cite{Lyons1998} and~\cite{gubinelli2010}, and in the theory of regularity structures introduced in~\cite{hairer2014} and further developed in \cite{BrunedARR19}. In numerical analysis, trees play a central role in the study of Butcher series and geometric numerical integration, starting from the work of~\cite{butcherCoefficientsStudyRungeKutta1963} and further developed in \cite{hairer1974, HairerWannerGNI, MuruaHAR06, ChartierASB10}. In these settings, one studies algebras defined on classes of graphs (most commonly rooted trees or forests) and analyzes the algebraic and combinatorial properties that emerge.

Working with such algebras quickly leads to symbolic computations that are intricate and repetitive. Despite their importance, general-purpose software for computations with algebras of trees and forests remains limited.
Existing implementations are often tailored to specific applications or implemented within larger computer algebra systems, where they are developed for narrow purposes and do not provide a unified or easily extensible framework for general computations with trees and forests.

In this paper, we present \texttt{Arboretum.hs}, a Haskell package that provides a systematic and rigorous framework for computations involving algebras of graphs. The package is designed with generality and composability as primary goals. It introduces basic abstractions for graded vector spaces and provides tools for defining and manipulating algebras of trees and forests in a way that closely follows their mathematical definitions.

As a concrete example, we implement the pre-Lie grafting algebra of decorated forests, which plays a role in the analysis of numerical integrators. The package also includes tools for generating graphical PDF representations of decorated forests, with support for customization and straightforward extension to more general classes of graphs.

The design of \texttt{Arboretum.hs} is guided by three core principles: 
\begin{enumerate}
    \item Readability: the code mirrors the underlying mathematics as closely as possible, making implementations easy to follow,
    \item Extensibility: new structures and operations can be added without major changes to the existing codebase,
    \item Testing: correctness is emphasized throughout, and the package is structured to make validation of algebraic properties straightforward.
\end{enumerate}

These goals strongly influenced the choice of Haskell as the implementation language. Haskell’s syntax and functional style are well suited to expressing algebraic definitions, particularly recursive objects such as trees and forests. Immutability and pure functions simplify reasoning about program behavior, while the static type system allows algebraic structures to be reflected directly in the code and checked at compile time. This stands in contrast to dynamically typed languages such as Python or Matlab, where similar errors often surface only at runtime.

Lazy evaluation is another important feature in this context. It allows computations to be deferred until their results are needed, which is particularly well suited to infinite objects such as Butcher series, which are formal sums of trees used in the analysis of numerical integrators and discussed in detail in Section~\ref{sec:butcher}. This enables a natural treatment of infinite sums without explicit truncation, an approach that is considerably more cumbersome in languages such as Python, Julia, or Matlab.

Haskell also emphasizes composability and modularity, making it easy to extend the codebase. In addition, property-based testing tools such as QuickCheck allow algebraic laws to be expressed directly as testable properties, providing a level of automated verification that is harder to achieve in dynamically typed environments.

We emphasize that \texttt{Arboretum.hs} prioritizes readability and user experience over raw performance. This choice is deliberate. The package is intended both as a computational tool and as a platform for exploration and experimentation with algebras of graphs. Where performance becomes critical, Haskell’s Foreign Function Interface allows integration with high-performance code written in C, Julia, or Python. To reduce the learning barrier, the package avoids advanced Haskell features such as monads or type-level programming, making it more approachable to users without extensive experience in functional programming.

Several related projects exist. A prominent example is \texttt{BSeries.jl}\footnote{\url{https://github.com/ranocha/BSeries.jl}, \cite{ranocha2021bseries}}, which replaces the earlier Python implementation \texttt{BSeries.py}\footnote{\url{https://github.com/ketch/BSeries}, \cite{BSeries.py}}. Its focus is on efficient implementations of classical Butcher series techniques and Runge-Kutta methods. Developed with different priorities, \texttt{BSeries.jl} primarily targets numerical analysts and well-established structures from numerical analysis. As a result, it is less flexible from an algebraic and combinatorial perspective, and less suited for exploring extensions of tree and forest algebras beyond the classical setting; see \cite{ketcheson2023computing} for details.

We also mention the master’s thesis of \cite{Sundklakk15pybs}, which introduces a Python package \texttt{pybs}\footnote{\url{https://github.com/henriksu/pybs}, \cite{pybs}} for automating classical computations in the theory of Butcher series. Additionally, \cite{munthe-kaas2018} present recursive formulas for operations on planar forests, motivated in part by the development of a Haskell package for their computation, and includes a complementary discussion of the benefits of using Haskell in this context.

Several Haskell packages provide complementary tools for algebraic and combinatorial computations. For example, \texttt{vector-space}\footnote{\url{https://hackage.haskell.org/package/vector-space}} offers generic abstractions for vector spaces and linear operations, \texttt{algebraic-graphs}\footnote{\url{https://hackage.haskell.org/package/algebraic-graphs}, \cite{MokhovAGC17}} supports compositional construction and manipulation of graphs, and \texttt{free-algebras}\footnote{\url{https://hackage.haskell.org/package/free-algebras}} enables symbolic computation with generic algebraic structures. While powerful in their respective domains, these libraries are either too general to capture the full combinatorial and algebraic structure of trees and forests, or they are not tailored for symbolic computations of operations such as grafting, Grossman–Larson products, or the Connes–Kreimer coproduct. Arboretum.hs addresses this gap by providing a practical, type-safe framework that serves as a flexible platform for research in algebraic combinatorics, rough paths, regularity structures, and numerical analysis.

Most of the work on the \texttt{Arboretum.hs} package was carried out during the first author’s PhD studies, which focused on the analysis of numerical integrators using algebraic and combinatorial tools, and a discussion of the package is included in first author's PhD thesis \cite{BronascoThesis}.

The paper is structured in the following way. Section \ref{sec:getting_started} contains the instructions on how to install and use the package. Section \ref{sec:butcher} introduces Butcher series, which are the main motivation for the develoment of the package and an important tool in the analysis of numerical integrators. Section \ref{sec:vector_space} presents the necessary tools to represent formal sums of trees and forests and Section \ref{sec:algebra} describes the implementation of the algebraic and coalgebraic structures over trees and forests.

\section{Getting started}
\label{sec:getting_started}

The reader is encouraged to explore the examples provided in this paper for a practical understanding of the package’s functionality. To begin using \texttt{Arboretum.hs}, install the Haskell Tool Stack (Stack) by following the instructions from
\begin{center} \href{https://haskellstack.org}{\texttt{https://haskellstack.org}}, \end{center}
clone the git repository
\begin{center} \href{https://gitlab.unige.ch/Eugen.Bronasco/arboretum.hs}{\texttt{https://gitlab.unige.ch/Eugen.Bronasco/arboretum.hs}} \end{center}
open a terminal in the root folder and run the command \texttt{stack repl}. The initial release version of the package is also availabe at \cite{bronasco_zenodo}.
The package uses a custom extension of the LaTeX package \texttt{planarforest} to draw the trees and forests. The original \texttt{planarforest} package is available at
\begin{center}\href{https://hmarthinsen.github.io/planarforest/}{\texttt{https://hmarthinsen.github.io/planarforest/}} \end{center}

 The \texttt{display} function is used to display vectors of trees and forests. It generates a PDF file in the root folder called \texttt{output.pdf} which contains the visual representation of the vector and attemps to open the file using the default document viewer. The \texttt{display} function is used in the examples below to illustrate the results of the computations.

\begin{codeexample}
  \label{uex:nonplanar_forests}
  Below you can find some examples of the package usage where we define two forests \texttt{f1} and \texttt{f2}, and graft \texttt{f1} onto \texttt{f2}. The details can be found in Section \ref{sec:grafting_of_trees}.
  \begin{haskellcode*}{}
>>> f1 = fromBrackets "1[2],1[2]" :: MultiSet (Tree Integer)
>>> display f1
  \end{haskellcode*}
  $ (\forestA \cdot \forestB) $ 
  \begin{haskellcode*}{}
>>> f2 = fromBrackets "1,1[2,2]" :: MultiSet (Tree Integer)
>>> display f2
  \end{haskellcode*}
  $ (\forestC \cdot \forestD) $ 
  \begin{haskellcode*}{}
>>> display $ graft f1 f2
  \end{haskellcode*}
 $ 2(\forestE \cdot \forestF) + 2(\forestG \cdot \forestH) + 4(\forestI \cdot \forestJ) + (\forestK \cdot \forestL) + 4(\forestM \cdot \forestN) + 2(\forestO \cdot \forestP) + (\forestQ \cdot \forestR) $ 
\end{codeexample}

\section{Butcher series}
\label{sec:butcher}

In this section, we introduce Butcher series, introduced in \cite{hairer1974} and named B-series to honor the seminal algebraic work of Butcher \cite{butcherCoefficientsStudyRungeKutta1963,Butcher_72}, see also \cite{HairerWannerGNI,ButcherBAA21}. They are a central tool in the convergence analysis of numerical integrators and in the study of their geometric properties. Butcher series form a rich mathematical framework that connects applied mathematics, in particular numerical analysis, with areas of pure mathematics such as combinatorics, combinatorial algebra, and geometry. Their formulation relies heavily on rooted non-planar trees and forests. Butcher series and their many extensions have been used in the analysis of numerical integrators for Hamiltonian systems on $\R^d$ often used in mechanics, (stochastic) ordinary differential equations on $\R^d$ and on manifolds as well as in the theory of (stochastic) partitioned differential equations. See \cite{connesHopfAlgebrasRenormalization1998,BrouderRKM00,HairerWannerGNI, MuruaHAR06, ChartierASB10, gubinelli2010, BrunedARR19}.

In this paper, we restrict our attention to the classical theory of Butcher series which is developed in the context of ordinary differential equations on $\R^d$.
Let us consider the following ordinary differential equation which describes the change of a position vector $y(t) \in \R^d$ in time:
\begin{equation}
    \label{eq:ODE}
    \frac{dy(t)}{dt} = f(y(t)) \,, \quad \text{with } y(0) = y_0 \in \R^d \,,
\end{equation}
for some given vector $y_0 \in \R^d$ where $f : \R^d \to \R^d$ is a smooth vector field which assigns to every position in $\R^d$ a vector whose direction and magnitude describe the direction and speed of the movement of $y(t)$ as time passes. A solution of \eqref{eq:ODE} is a function $y(t)$ such that the equation \eqref{eq:ODE} is satisfied. It is often the case that an explicit formula for the solution $y(t)$ cannot be found, and, therefore, the only possible way to compute $y(t)$ is by approximating its value using a numerical integrator for $t \in [0, T]$ for some final time $T \in \R$.

Let $y_n$ be an approximation of the solution $y(t_n)$ at time $t_n = n h$, where $h > 0$ is the timestep size. 
Runge-Kutta methods are a widely used class of numerical integrators which include
\begin{align*}
    \text{Forward Euler:} \quad & y_{n+1} = y_n + h f(y_n) \,, \\
    \text{Backward Euler:} \quad & y_{n+1} = y_n + h f(y_{n+1}) \,, \\
    \text{Implicit midpoint:} \quad & y_{n+1} = y_n + h f \big(\frac{1}{2} (y_n + y_{n+1})\big) \,, \\
    \text{Trapezoidal method:} \quad & y_{n+1} = y_n + \frac{h}{2} f (y_n) + \frac{h}{2} f(y_{n+1}) \,,
\end{align*}
as well as the commonly used RK4 method. 
The idea behind these methods is to compute successive approximations $y_0, y_1, \ldots, y_N$ of the solution at times $t_0, t_1, \ldots, t_N$ with $N = T/h$.

One of the key characteristics of a numerical integrator is its \emph{order of convergence} where a numerical integrator $y_1 := \Phi_h (y_0)$ has order $p$ if 
\[ | y(T) - \Phi_h^N (y_0) | \approx \OO(h^p) \,. \]
Order of convergence indicates how quickly the numerical solution converges to the exact solution as the timestep size $h$ decreases.
For example, forward and backward Euler methods have order $1$, implicit midpoint and trapezoidal have order $2$, and the RK4 method has order $4$.

The order of convergence of Runge–Kutta methods is analyzed using Butcher series, whose central idea is to represent the Taylor expansions of both the exact and numerical solutions as formal sums indexed by rooted, non-planar trees.
A commutative monomial $\tau_1 \cdots \tau_n$ for some $n \in \N$ is a product of trees $\tau_1, \dots, \tau_n$ in which the order of trees does not matter.

\begin{definition}
    \label{def:tree}
    A \emph{rooted non-planar tree} $\tau$ is a tuple $(r, \pi)$ of a vertex $r$ called the \emph{root} of $\tau$ and a commutative monomial of rooted non-planar trees $\pi = \tau_1 \cdots \tau_n$ called the \emph{branches} of $\tau$. 
\end{definition}

The set of rooted non-planar trees is denoted by $T$ and the commutative monomials of the rooted non-planar trees are called \emph{rooted non-planar forests} and form the set $F$. We note that all trees and forests are assumed to be rooted and non-planar and, therefore, we omit writting it from now on. All trees with up to $4$ vertices can be found below with roots displayed at the bottom,
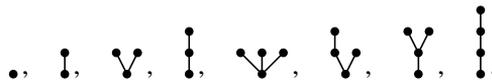
\[ \forestS\,,\quad \forestT\,,\quad \forestU \,, \quad \forestV \,, \quad \forestW \,, \quad \forestX \,, \quad \forestY \,, \quad \forestAB \,. \]
We note that the order of branches does not matter, that is, $\forestBB = \forestCB$. Some examples of forests can be found below,
\[ \one \,, \qquad \forestDB \,, \qquad \forestEB \,, \qquad \forestFB \,, \]
where $\one$ denotes the empty forest. We note that $T \subset F$ and that the order of trees does not matter, that is, $\forestGB = \forestHB$.

The sets $T$ and $F$ are isomorphic through the map $B^+_v : F \to T$ which constructs a tree $B^+_v(\pi)$ with root $v$ and branches $\pi$, for example,
\[ B^+_\bullet(\one) = \forestIB \,, \quad B^+_\bullet(\forestJB) = \forestKB \,, \quad B^+_\bullet(\forestLB) = \forestMB \,. \]

\cite{Cayley1857} describes a correspondence between trees and \emph{elementary differentials} which are the terms appearing in Taylor expansions of the exact and numerical solutions. This correspondence is defined in Definition \ref{def:F}.

\begin{definition}
    \label{def:F}
    Let $\F$ denote the corresondence between trees and elementary differentials defined as
    \[ \F (B^+_\bullet (\tau_1 \cdots \tau_n)) := h \sum_{i_1, \dots, i_n = 1}^d \F(\tau_1)^{i_1} \cdots \F(\tau_n)^{i_n} \partial_{i_1, \dots, i_n} f \,, \]
    where $v^i$ denotes $i^{th}$ component of a vector $v$, $\partial_i$ denotes the partial derivative with respect to the $i^{th}$ argument, and $\partial_{i_1, \dots, i_n} := \partial_{i_1} \cdots \partial_{i_n}$.
\end{definition}

For example,
\[ \F(\forestNB) = hf \,, \quad \F(\forestOB) = h^2 \sum_{i=1}^d f^i \partial_i f\,, \quad \F(\forestPB) = h^4 \sum_{i,j,k=1}^d f^i (\partial_i f^j) f^k (\partial_{j,k} f) \,. \]

Using an alternative notation, the values of $\F$ can be written as
\[ \F (B^+_\bullet (\tau_1 \cdots \tau_n)) = hf^{(n)} \big( \F(\tau_1), \dots, \F(\tau_n) \big) \,, \]
which is the directional derivative of $hf$ in the directions $\F(\tau_1), \dots, \F(\tau_n)$.

Taylor expansion of both exact and numerical solutions $y(h)$ and $y_1 = \Phi_h(y_0)$ given by Runge-Kutta methods can be expressed using Butcher series denoted by $B(1/\gamma, y_0)$ and $B(a, y_0)$ for sufficiently small timestep $h$,
\begin{align}
    y(h) &= y_0 + \sum_{\tau \in T} \frac{1}{\gamma(\tau) \sigma(\tau)} \F (\tau)(y_0) =: y_0 + B (1/\gamma, y_0) \,, \label{eq:exact_Bseries} \\
    \Phi_h (y_0) &= y_0 + \sum_{\tau \in T} \frac{a(\tau)}{\sigma(\tau)} \F (\tau)(y_0) =: y_0 + B (a, y_0) \,, \label{eq:numerical_Bseries}
\end{align}
where the coefficient map $\sigma : T \to \R$ denotes the symmetry of a tree and is defined as
\begin{equation}
    \label{eq:symmetry_coeff}
    \sigma \big(B^+_v(\tau_1^{k_1} \cdots \tau_n^{k_n})\big) := \prod_{i=1}^n k_i! \sigma(\tau_i) \,, 
\end{equation}
where $\tau_1, \dots, \tau_n$ are distinct and $k_1, \dots, k_n$ are the multiplicities of the corresponding trees while the coefficient map $\gamma : T \to \R$ denotes the factorial of a tree and is defined as
\[ \gamma(\tau) = \gamma \big(B^+_v (\tau_1 \cdots \tau_n)\big) = |\tau| \big( \prod_{i=1}^n \gamma(\tau_1) \big) \,, \]
where $|\tau|$ denotes the number of vertices of the tree $\tau$,
\[ | B^+_v (\tau_1 \cdots \tau_n) | :=  1 + \sum_{i=1}^n |\tau_i| \,. \]
Coefficient map $a : T \to \R$ depends on the particular choice of the Runge-Kutta method, see \cite{HairerSOD93} for details. Values of the coefficient maps $\sigma, \gamma : T \to \R$ for all trees up to size $4$ can be found in Table \ref{tab:coeff_maps}

\begin{table}[h]
    \centering
    \begin{tabular}{c|c|c|c|c|c|c|c|c}
        Tree $\tau$ & $\forestQB$ & $\forestRB$ & $\forestSB$ & $\forestTB$ & $\forestUB$ & $\forestVB$ & $\forestWB$ & $\forestXB$ \\ \hline
        $\sigma (\tau)$ & $1$ & $1$ & $2$ & $1$ & $6$ & $2$ & $2$ & $1$ \\ \hline
        $\gamma (\tau)$ & $1$ & $2$ & $3$ & $6$ & $4$ & $8$ & $12$ & $24$
    \end{tabular}
    \caption{Symmetry $\sigma(\tau)$ and the factorial $\gamma(\tau)$ for all trees $\tau$ of with up to $4$ vertices.}
    \label{tab:coeff_maps}
\end{table}

The description of the exact and numerical solutions \eqref{eq:exact_Bseries} and \eqref{eq:numerical_Bseries} in terms of Butcher series gives a straightforward way to derive order conditions for Runge-Kutta methods. Indeed, a Runge-Kutta method has order $p$ if and only if the corresponding coefficient map $a : T \to \R$ satisfies $a(\tau) = 1/\gamma(\tau)$ for all trees $\tau$ of size $|\tau| \leq p$. This is a major milestone in the study of Runge-Kutta methods obtained in \cite{butcherCoefficientsStudyRungeKutta1963} and is the original motivation for the introduction of Butcher series.

\subsection{Algebras of forests and composition of Butcher series}
\label{sec:butcher:algebra}

Let $y_1 = \Phi_h(y_0)$ and $y_1 = \Psi_h(y_0)$ be two numerical integrators that can be Taylor expanded as Butcher series $B(a,y_0)$ and $B(b, y_0)$, respectively. We compose the integrators $\Phi_h$ and $\Psi_h$ by first applying $\Phi_h$ to $y_0$ and then applying $\Psi_h$ to the result, that is, we compute $\Psi_h \big( \Phi_h (y_0) \big)$.
In this section, we describe the Butcher series corresponding to the composition,
\[ \Psi_h \big( \Phi_h (y_0) \big) = B(b, B(a, y_0)) = B(a * b, y_0) \,, \]
where $a * b : T \to \R$ defines a new coefficient map which is described using the algebraic structures introduced in this section, which include: grafting algebras over trees and forests, symmetric algebra over forests, and combinatorial Hopf algebras of Grossman-Larson and Connes-Kreimer. Implementations of these algebras are discussed in Section \ref{sec:algebra}.

Elementary differentials are vector fields $g : \R^d \to \R^d$ that appear in the Taylor expansion of the exact solution of \eqref{eq:ODE}. For each elementary differential $g : \R^d \to \R^d$, there exists a tree $\tau \in T$ such that $\dF(\tau) = g$, see Definition \ref{def:F}.
The set of elementary differentials is denoted by $\dF (T) := \{ \dF(\tau) \; | \; \tau \in T \}$. Let $\TT$ denote the vector space with the set of trees $T$ as basis, then let $\dF(\TT)$ be the vector space with the set of vector fields $\dF(T)$ as basis. 

\begin{definition}
    \label{def:vec_space}
    A \emph{vector space} is a set $V$ and a field of scalars $k$ with two operations: addition $+$ and scalar multiplication $\cdot$ that satisfy the following properties:
    \begin{enumerate}
        \item[(1)] $V$ is an abelian group under addition,
        \item[(2)] for all $a, b \in V$ and $\lambda \in k$, we have $ \lambda \cdot (a + b) = \lambda \cdot a + \lambda \cdot b$,
        \item[(3)] for all $a \in V$ and $\lambda, \mu \in k$, we have $(\lambda + \mu) \cdot a = \lambda \cdot a + \mu \cdot a$ and $(\lambda \mu) \cdot a = \lambda \cdot (\mu \cdot a)$,
        \item[(4)] for all $a \in V$, we have $1 \cdot a = a$.
    \end{enumerate}
    A set $\{ v_i \}_{i=1}^d$, $v_i \in V$, is called a \emph{basis} if for each $v \in V$ there exists a unique set of coefficients $\{ c_i \}_{i=1}^d$ with $c_i \in k$ such that 
    \[ v = c_1 v_1 + \cdots + c_d v_d \,. \]
    Vector space $V$ has \emph{dimension} $d$.
\end{definition}

The map $\dF$ between trees and elementary differentials (Definition \ref{def:F}), relates Butcher series and formal sums of trees,
\[ \sum_{\tau \in T} \frac{a(\tau)}{\sigma(\tau)} \tau \quad \overset{\dF}{\longmapsto} \quad \sum_{\tau \in T} \frac{a(\tau)}{\sigma(\tau)} \dF(\tau) = B(a, y_0) \,. \]
This allows us to replace elementary differentials by the concise and intuitive formalism of trees. This powerful formalism allows us to use the techniques from combinatorics and combinatorial algebra to study Butcher series. We denote the vector space of formal sums of trees by $\overline{T}$. 

\begin{definition}
    \label{def:graded_vec_space}
    A vector space $V$ is \emph{graded} if $V = \bigoplus_{n \in \mathbb{N}} V_n$, that is, each $v \in V$ can be decomposed uniquely as
    \[ v = \sum_{n \in \N} v_n \,, \quad v_n \in V_n \,, \]
    where each $V_n \subset V$ is a vector subspace and elements of $V_n$ are called \emph{homogeneous} elements of degree $n$.
\end{definition}

The vector space of trees $\TT$ and of formal sums of trees $\overline{\TT}$ are graded vector spaces with the grading given by the number of vertices. For example,
\begin{align*}
    \overline{\TT}_1 &= \TT_1 := \text{span}\{ \forestYB \} \,, &\quad \overline{\TT}_2 &= \TT_2 := \text{span}\{ \forestAC \} \,, \\
    \overline{\TT}_3 &= \TT_3 := \text{span}\{\forestBC, \forestCC\} \,, &\quad \overline{\TT}_4 &= \TT_4 := \text{span}\{ \forestDC, \forestEC, \forestFC, \forestGC \} \,.
\end{align*}
We note that each $\overline{\TT}_n$ for $n \in \N$ is finite-dimensional. The implementation of the infinite-dimensional graded vector spaces whose homogeneous subspaces are finite-dimensional is discussed in Section \ref{sec:vector_space}.

\subsubsection{Grafting algebra of trees}

Let $\{e_i\}_{i=1}^d$ be the canonical basis of $\R^d$, that is,
\[ e_1 = (1,0,\dots,0)^T \,, \quad e_2 = (0,1,0,\dots,0)^T \,, \quad \dots \,, \quad e_d = (0, \dots, 0, 1)^T \,. \]
Then, a vector field $g : \R^d \to \R^d$ can be written as $g(x) = \sum_{i=1}^d g^i(x) e_i$ for $g^i : \R^d \to \R$. We identify the space of vector fields with the space of first-order differential operators by identifying  $\{ e_i \}_{i=1}^d$ with $\{ \partial_i \}_{i=1}^d$. Given a first-order differential operator $g : \R^d \to \R^d$, its application to $h : \R^d \to \R^d$ is given by
\[ g[h] (x) = \sum_{i=1}^d g^i (x) (\partial_i h) (x) = \sum_{i,j=1}^d g^i (x) (\partial_i h^j) (x) \partial_j \,. \]
We omit writting $x$ from now on for conciseness. The space of first-order differential operators $\dF(\TT)$ forms a pre-Lie algebra $(\dF(\TT), \blank [\blank])$, that is, for any $g_1, g_2, g_3 \in \dF(\TT)$, we have the following relation,
\[ g_1 [g_2[g_3]] - g_1[g_2][g_3] = g_2 [g_1[g_3]] - g_2[g_1][g_3] \,. \]

Let us define the grafting product $\graft : \TT \otimes \TT \to \TT$ such that 
\[ \dF(\tau \graft \gamma) = \dF(\tau)[\dF(\gamma)] \,. \]

\begin{definition}
    \label{def:grafting}
    Let $\tau, \gamma \in T$ and let $V(\gamma)$ be the set of vertices of $\gamma$. The \emph{grafting product} is defined as
    \[ \tau \graft \gamma := \sum_{v \in V(\gamma)} \tau \graft_v \gamma \,, \]
    where $\tau \graft_v \gamma$ is the tree obtained by attaching the root of $\tau$ to the vertex $v$ of $\gamma$.
\end{definition}

For example,
\[ \forestHC \graft \forestIC = \forestJC + \forestKC \,, \quad \forestLC \graft \forestMC = \forestNC + \forestOC + \forestPC + \forestQC \,. \]

The algebra $(\TT, \graft)$ is the free pre-Lie algebra with one generator $\{\bullet\}$ as is shown in \cite{chapotonPreLieAlgebrasRooted2001}, that is, the relation
\[ \tau_1 \graft (\tau_2 \graft \tau_3) - (\tau_1 \graft \tau_2) \graft \tau_3 = \tau_2 \graft (\tau_1 \graft \tau_3) - (\tau_2 \graft \tau_1) \graft \tau_3 \,, \quad \tau_1, \tau_2, \tau_3 \in \TT \,, \]
is the only relation satisfied by the product $\graft : \TT \otimes \TT \to \TT$ and each $\tau \in \TT$ can be constructed using only the generator $\bullet$ and the grafting product $\graft$. For example,
\begin{align*}
    \forestRC &= \bullet \graft \bullet \,, &\quad \forestSC &= \forestTC \graft \forestUC - \forestVC \,, \\
    \forestWC &= \forestXC \graft \bullet \,, &\quad \forestYC &= \forestAD \graft \bullet \,, \\
    \forestBD &= \bullet \graft \forestCD - \forestDD \,, &\quad \forestED &= \bullet \graft \forestFD - 2 \forestGD \,. \\
    \forestHD &= \forestID \graft \bullet \,,
\end{align*}
The implementation of $(\TT, \graft)$ is discussed in Section \ref{sec:grafting_of_trees}.
The grafting product $\graft$ over $\TT$ is extended to $\overline{\TT}$. Moreover, it respects the grading, that is,
\[ \TT_n \graft \TT_m \subset \TT_{n+m} \,. \]
Such products are called \emph{graded} and the algebras they form are called \emph{graded algebras}.

We note that using the chain rule of differentiation and Definition \ref{def:F} of $\dF$, we can express the $n^{th}$ derivative of the exact solution $y(t)$ of \eqref{eq:ODE} at $t=0$ scaled by $h^n$ as
\[ h^n y^{(n)}(0) = \dF \Big( \underbrace{\forestJD \graft \big(\forestKD \graft \cdots (\forestLD \graft \forestMD)\cdots\big)}_{n\text{ times}}\Big) (y_0) = \dF\big( \bullet^{\graft n} \big) (y_0) \,, \]
where $\bullet^{\graft n}$ is the grafting product of $\bullet$ taken $n$ times assuming right associativity.
This allows us to express the Taylor expansion of the exact solution of \eqref{eq:ODE} as
\[ y(h) = y_0 + \dF\big(\sum_{n=1} \frac{1}{n!} \bullet^{\graft n} \big)(y_0) = \dF \big( \exp^\graft (\bullet) \big) (y_0) \,, \]
where $\exp^\graft$ denotes the exponential with respect to the grafting product.
We obtain the Butcher series representation of the exact solution \eqref{eq:exact_Bseries} by grouping the terms.

\subsubsection{Symmetric algebra of forests}
\label{sec:butcher:symmetric_algebra}

In this subsection, we construct the algebra of high-order differential operators which extends the algebra $(\dF(\TT), \blank [\blank])$ of first-order differential operators.
Let $V$ be a vector space spanned by $\{v_i\}_{i \in I}$ for a countable set of indices $I$ and let $S(V)$ be the vector space spanned by commutative monomials $v_1 \cdots v_n$ for $n \in \N$ where the order of $v_1, \dots, v_n$ does not matter. The space $S(V)$ is a graded algebra with the concatenation product,
\[ (v_1 \cdots v_n) \cdot (u_1 \cdots u_m) = v_1 \cdots v_n u_1 \cdots u_m \,, \]
and the empty monomial $\one$ as the unit. The grading is given by the number of the elements of $V$ in a monomial, for example, the grading of $v_1 \cdots v_n$ is $n$.
Let us define the deshuffle coproduct $\Delta : S(V) \to S(V) \otimes S(V)$ over the vector space $S(V)$.
\begin{definition}
    \label{def:deshuffle}
    Let the \emph{deshuffle coproduct} $\Delta : S(V) \to S(V) \otimes S(V)$ be defined as
    \begin{align*}
        \Delta (\one) &= \one \otimes \one \,, \\
        \Delta (v) &= v \otimes \one + \one \otimes v \,, \quad v \in V \,, \\
        \Delta (a \cdot b) &= \Delta(a) \cdot \Delta (b) \,, \quad a \in S(V) \,,
    \end{align*}
    where $(a_1 \otimes b_1) \cdot  (a_2 \otimes b_2) = (a_1 \cdot a_2) \otimes (b_1 \cdot b_2)$ for $a_1, a_2, b_1, b_2 \in S(V)$.
\end{definition}
Intuitively, the deshuffle coproduct sums over all ways to place the elements of $V$ on the left or the right side of the tensor product while preserving their order. For example,
\begin{align*}
    \Delta (v_1 v_2 v_3) = v_1 v_2 v_3 &\otimes \one + v_2 v_3 \otimes v_1 + v_1 v_3 \otimes v_2 + v_1 v_2 \otimes v_3 \\
    + v_3 &\otimes v_1 v_2 + v_2 \otimes v_1 v_3 + v_1 \otimes v_2 v_3 + \one \otimes v_1 v_2 v_3 \,.
\end{align*}

The space $S(V)$ is a coalgebra with the deshuffle coproduct $\Delta : S(V) \to S(V) \otimes S(V)$ and counit $\delta_\one : S(V) \to \R$ defined as
\[ \delta_\one (a) := \begin{cases} 1 \,, \quad a = \one \,, \\ 0 \,, \quad a \neq \one \,. \end{cases} \]
The deshuffle coalgebra is graded, that is,
\[ \Delta(S(V)_n) \subset \sum_{i=0}^n S(V)_i \otimes S(V)_{n-i} \,, \quad \delta_\one (S(V)_{n \geq 1}) = \{0\} \,. \]

The space $S(V)$ is a combinatorial Hopf algebra $(S(V), \one, \cdot, \delta_\one, \Delta)$ called the \emph{symmetric algebra}, see Definition \ref{def:comb_hopf_alg}. The implementation of the symmetric algebra $S(V)$ is discussed in Section \ref{sec:free_algebras}.

\begin{definition}
    \label{def:comb_hopf_alg}
    A graded vector space $V$ with $\text{dim}(V_0) = 1$ endowed with a graded algebra $(V, \cdot)$ with a unit $\one$ and coalgebra $(V, \Delta)$ with counit $\epsilon : V \to \R$ such that
    \begin{enumerate}
        \item $\epsilon (\one) = 1$,
        \item $\Delta(\one) = \one \otimes \one$,
        \item $\epsilon (a \cdot b) = \epsilon (a) \cdot \epsilon (b)$,
        \item $\Delta(a \cdot b) = \Delta(a) \cdot \Delta(b)$,
    \end{enumerate}
    is referred to as a \emph{combinatorial Hopf algebra} $(V, \one, \cdot, \epsilon, \Delta)$. We recall that $(a_1 \otimes b_1) \cdot (a_2 \otimes b_2) = (a_1 \cdot a_2) \otimes (b_1 \cdot b_2)$.
\end{definition}

Given a pre-Lie product on a vector space $V$, \cite{oudomLieEnvelopingAlgebra2008} extend the pre-Lie product to the symmetric algebra $S(V)$. In the case of $V = \TT$, this corresponds to the extention of the grafting product $\graft$ to the space of forests $S(\TT) = \FF$ spanned by the set of forests $F$. The grafting product $\graft : \FF \otimes \FF \to \FF$ is defined, for $\tau \in \TT$ and $\pi, \eta, \mu \in \FF$, as
\begin{align}
    \one \graft \one &= \one \,, \quad \tau \graft \one = 0 \,, \label{eq:GuinOudom} \\
    (\tau \cdot \pi) \graft \eta &= \tau \graft (\pi \graft \eta) - (\tau \graft \pi) \graft \eta \,, \nonumber \\
    \pi \graft (\eta \cdot \mu) &= \sum_{(\pi)} (\pi_{(1)} \graft \eta) \cdot (\pi_{(2)} \graft \mu) \,, \nonumber
\end{align}
where we use the notation $\Delta (\pi) = \sum_{(\pi)} \pi_{(1)} \otimes \pi_{(2)}$ for the deshuffle coproduct of $\pi$, see Definition \ref{def:deshuffle}. 
For example,
\begin{align*}
    \forestND \graft \forestOD &= 2 \forestPD + 2 \forestQD \,, \\
    \forestRD \graft \forestSD &= \forestTD + \forestUD + \forestVD + \forestWD \,.
\end{align*}
Intuitively, the grafting product $\graft : \FF \otimes \FF \to \FF$ sums over all ways to graft each tree of the forest on the left to each vertex of the forest on the right.

The algebra of high-order differential operators $(\dF(\FF), \blank [\blank])$ where $\dF(\FF)$ has $\dF(F) := \{ \dF(\pi) \; | \; \pi \in F \}$ as basis satisfies
\[  \dF(\pi \graft \mu) = \dF(\pi) [\dF(\mu)] \,. \]
We note that $\dF(B^+(\pi)) = \dF(\pi \graft \bullet) = h \dF(\pi)[f]$ where $\pi \in F$. This gives an alternative and purely algebraic definition to the map $\dF$ which can be seen as the unique map which maps the generator $\bullet$ to $hf$ and the grafting product $\graft$ to $\blank [\blank]$.

The grafting product $\graft : \FF \otimes \FF \to \FF$ arises naturally in the Taylor expansion of $\dF(\tau)(y_0 + \dF(\gamma))$ for $\tau, \gamma \in \TT$, in particular,
\begin{equation}
    \label{eq:concat_expansion}
    \dF(\tau)(y_0 + \dF(\gamma)) = \dF \big( \exp^\cdot(\gamma) \graft \tau \big)(y_0) \,,
\end{equation}
where $\exp^\cdot$ denotes the exponential with respect to the concatenation product. The property \eqref{eq:concat_expansion} is an essential step in the expansion of Runge-Kutta methods as Butcher series \eqref{eq:numerical_Bseries}.

\subsubsection{Grossman-Larson and Connes-Kreimer Hopf algebras of forests}

The grafting algebra $(\FF, \graft)$ is used to study the algebra of differential operators $(\dF(\FF), \blank [\blank])$ where the product $\dF(\pi) [\dF(\eta)]$ is the differentiation of the differential operator $\dF(\eta)$ by the differential operator $\dF(\pi)$.

Another natural algebraic structure on $\dF(\FF)$ consists in the composition of differential operators, that is,
\begin{equation}
    \label{eq:comp_diff_op}
    \dF(\pi)\big[ \dF(\eta) [\phi] \big] = \big( \dF(\pi) \circ \dF(\eta) \big) [\phi] \,,
\end{equation}
where $\phi$ can be a function $\R^d \to \R$, a vector field $\R^d \to \R^d$, or a differential operator.
This defines an algebra $(\dF(\FF), \circ)$ of differential operators. \cite{oudomLieEnvelopingAlgebra2008} introduce the corresponding algebra over $\FF$ which is called the Grossman-Larson algebra with the Grossman-Larson product defined, for $\pi, \eta \in \FF$, as
\[ \pi \gl \eta := \sum_{(\pi)} \pi_{(1)} \cdot (\pi_{(2)} \graft \eta) \,. \]
Some examples of performing the Grossman-Larson can be found below,
\begin{align*}
    \forestXD \gl \forestYD &= \forestAE + \forestBE + \forestCE \,, \\
    \forestDE \gl \forestEE &= \forestFE + \forestGE + \forestHE + \forestIE + \forestJE + \forestKE + \forestLE + \forestME + \forestNE \,.
\end{align*}
Intuitively, the Grossman-Larson product $\gl : \FF \otimes \FF \to \FF$ sums over all ways to split the trees of the forest on the left into two groups: the first group is concatenated to the forest on the right, and the second is grafted onto the forest on the right.
Grossman-Larson product satisfies, for $\pi, \eta, \mu \in \FF$,
\[ \pi \graft (\eta \graft \mu) = (\pi \gl \eta) \graft \mu \,, \]
which implies $\dF(\pi \gl \eta) = \dF(\pi) \circ \dF(\eta)$ using \eqref{eq:comp_diff_op} and gives an alternative definition of the Grossman-Larson product,
\[ \pi \gl \eta = B^-_\bullet \big( \pi \graft B^+_\bullet (\eta) \big) \,, \quad \text{for } \pi, \eta \in \FF \,, \]
where $B^-_\bullet$ removes the root $v$. Recall that $B^+_\bullet (\eta) = \eta \graft \bullet$. 
Grossman-Larson algebra has the empty forsest as the unit and, together with the deshuffle coalgebra, forms the Grossman-Larson combinatorial Hopf algebra $(\FF, \gl, \one, \Delta, \delta_\one)$ graded by the number of vertices. The implementation of the Grossman-Larson Hopf algebra is discussed in Section \ref{sec:grafting_of_trees}.

Let us define the Connes-Kreimer coproduct $\Delta_{CK} : \FF \to \FF \otimes \FF$.
\begin{definition}
    \label{def:ck}
    \emph{Connes-Kreimer coproduct} $\Delta_{CK} : \FF \to \FF \otimes \FF$ is defined, for $\pi, \eta \in F$, as
    \begin{align*}
        \Delta_{CK} (B^+_v (\pi)) &:= B^+_v (\pi) \otimes \one + \sum_{(\pi)} \pi_{(1)} \otimes B^+_v (\pi_{(2)}) \,, \\
        \Delta_{CK} (\pi \cdot \eta) &:= \Delta_{CK} (\pi) \cdot \Delta_{CK}(\eta) \,,
    \end{align*}
    and $\Delta_{CK} (\one) = \one \otimes \one$ where we use the notation $\Delta_{CK}(\pi) = \sum_{(\pi)} \pi_{(1)} \otimes \pi_{(2)}$ and
    \[ (a_1 \otimes b_1) \cdot (a_2 \otimes b_2) = (a_1 \cdot a_2) \otimes (b_1 \cdot b_2) \,. \]
\end{definition}

For example,
\begin{align*}
    \Delta_{CK} (\forestOE) &= \forestPE \otimes \one + \one \otimes \forestQE \,, \\
    \Delta_{CK} (\forestRE) &= \forestSE \otimes \one + \forestTE \otimes \forestUE + \one \otimes \forestVE \,, \\
    \Delta_{CK} (\forestWE) &= \forestXE \otimes \one  + \forestYE \otimes \forestAF + \forestBF \otimes \forestCF + \one \otimes \forestDF \,, \\
    \Delta_{CK} (\forestEF) &= \forestFF \otimes \one + \forestGF \otimes \forestHF + 2 \forestIF \otimes \forestJF + \one \otimes \forestKF \,, \\
    \Delta_{CK} (\forestLF) &= \forestMF \otimes \one + \forestNF \otimes \forestOF + \forestPF \otimes \forestQF + \forestRF \otimes \forestSF + \forestTF \otimes \forestUF + \forestVF \otimes \forestWF + \one \otimes \forestXF \,.
\end{align*}

The vector space $\FF$ of forests together with the concatenation product, unit $\one$, and Connes-Kreimer coproduct with the counit $\delta_\one$, forms the Connes-Kreimer combinatorial Hopf algebra $(\FF, \one, \cdot, \delta_\one, \Delta_{CK})$.

We consider the inner product $\langle \blank, \blank \rangle_\sigma : \FF \otimes \FF \to \R$ defined, for $\pi, \eta \in \FF$ as
\[ \langle \pi, \eta \rangle_\sigma := \begin{cases} \sigma (\pi) \,, \quad \text{if } \pi = \eta \,, \\ 0 \,, \quad \text{otherwise,} \end{cases} \]
where $\sigma (\pi)$ denotes the symmetry coefficient of the forest $\pi$, see \eqref{eq:symmetry_coeff}.
We rely on the following relations between the concatenation product and deshuffle coproduct, and Grossman-Larson product and Connes-Kreimer coproduct,
\begin{equation}
    \label{eq:adjoint_relations}
    \langle \pi \cdot \eta, \mu \rangle_\sigma = \langle \pi \otimes \eta, \Delta(\mu) \rangle_\sigma \,, \quad \langle \pi \gl \eta, \mu \rangle_\sigma = \langle \pi \otimes \eta, \Delta_{CK}(\mu) \rangle_\sigma \,,
\end{equation}
for all $\pi, \eta, \mu \in \FF$. In other words, concatenation and Grossman-Larson products are adjoints of the deshuffle and Connes-Kreimer coproducts, respectively, with respect to the inner product $\langle \blank, \blank \rangle_\sigma$.

Intuitively, this means that the coefficient of a forest $\mu \in F$ in $\pi \gl \eta$ is equal to the coefficient of $\pi \otimes \eta$ in $\Delta_{CK} (\mu)$ multiplied by the symmetry ratio $\sigma(\pi) \sigma(\eta) / \sigma(\mu)$. For example, the coefficient of $\forestYF$ in $\forestAG \gl \forestBG$ is $1$, which is equal to the coefficient of $\forestCG \otimes \forestDG$ in $\Delta_{CK} (\forestEG)$ multiplied by the symmetry ratio $1/2$.

The relation between Grossman-Larson and Connes-Kreimer Hopf algebras together with the properties of the Grossman-Larson product, \eqref{eq:concat_expansion}, \eqref{eq:adjoint_relations}, and the property
\[ a(\pi) = \langle \sum_{\eta \in F} \frac{a(\eta)}{\sigma(\eta)} \eta, \pi \rangle_\sigma \]
are enough to prove that the composition of two numerical integrators $B(a,y_0)$ and $B(b, y_0)$ results in a numerical integrator $B(a * b, y_0)$,
\begin{equation}
    \label{eq:composition_Bseries}
    B(b, B(a, y_0)) = B(a * b, y_0) \,, \quad \text{with } (a * b) (\tau) = \sum_{(\tau)} a(\tau_{(1)}) b(\tau_{(2)}) \,,
\end{equation}
where $\Delta_{CK} (\tau) = \sum_{(\tau)} \tau_{(1)} \otimes \tau_{(2)}$. We note that both $a, b : T \to \R$ are coefficient maps defined over trees, while in \eqref{eq:composition_Bseries} they are applied to forests $\tau_{(1)}, \tau_{(2)} \in F$, see the type signature of $\Delta_{CK}$. We extend the coefficient maps $a, b$ to forests by $a(\pi \cdot \eta) = a(\pi) a(\eta)$ and $a(\one) = b(\one) = 1$. Some values for the coefficient map $a * b$ can be found below,
\begin{align*}
    (a * b) (\forestFG) &= a(\forestGG) + b(\forestHG) \,, \\
    (a * b) (\forestIG) &= a(\forestJG) + a(\forestKG) b(\forestLG) + b(\forestMG) \,, \\
    (a * b) (\forestNG) &= a(\forestOG) + a(\forestPG) b(\forestQG) + a(\forestRG) b(\forestSG) + b(\forestTG) \,, \\
    (a * b) (\forestUG) &= a(\forestVG) + a(\forestWG)^2 b(\forestXG) + 2 a(\forestYG) b(\forestAH) + b(\forestBH) \,.
\end{align*}
The implementation of the Connes-Kreimer Hopf algebra is discussed in Section \ref{sec:CK}.

\subsection{Extensions of Butcher series}

We have presented a concise introduction to the theory of classical Butcher series for the analysis of numerical integrators of ordinary differential equations. From a computational point of view, this is the simplest setting, involving only basic algebraic structures and rooted non-planar trees.

More refined questions in numerical analysis quickly lead to substantially richer variants of the Butcher series formalism. In particular, long time and geometric analysis of numerical integrators gives rise to the Calaque--Ebrahimi-Fard--Manchon Hopf algebra introduced in \cite{calaqueTwoInteractingHopf2011}, whose coproduct poses serious computational challenges already in the simplest case that we consider here.

When Butcher series are applied to Hamiltonian systems found in mechanics, one is led to series indexed by non-rooted trees, where we forget which vertex is the root. See \emph{Chapter IX.9.2} in \cite{HairerWannerGNI} and \cite{bogfjellmoHamiltonianBseriesLie2017}. For example,
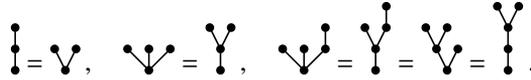
\[
    \forestCH = \forestDH \,, \quad \forestEH = \forestFH \,, \quad \forestGH = \forestHH = \forestIH = \forestJH \,.
\]

The extension of the Butcher series formalism to ordinary differential equations on manifolds leads to Lie-Butcher series and to Hopf algebras defined on planar trees and forests, where the order of branches in a tree and the order of trees in a forest matter. For instance,
\[
\forestKH \neq \forestLH
\quad \text{and} \quad
\forestMH \neq \forestNH .
\]
See \cite{Munthe-KaasLTR95} and \cite{munthe-kaasHopfAlgebraicStructure2006}.

In the stochastic setting, the theory of Butcher series introduces bicolored trees and forests, where one color represents noise, as well as exotic trees and forests in which vertices of a given color are paired. Examples of grafted trees include
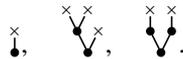
\[
\forestOH , \quad
\forestPH , \quad
\forestQH .
\]
Examples of exotic trees are
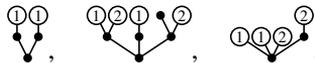
\[
\forestRH , \quad
\forestSH , \quad
\forestTH ,
\]
where the numbering of white vertices encodes the pairing, and the specific choice of labels is irrelevant. For example,
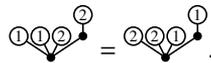
\[
\forestUH = \forestVH .
\]
See \cite{BurrageBurrage, RosslerRTA06, LaurentEAB20}.

Additional geometric constraints on numerical integrators, such as volume preservation, lead to the introduction of aromatic Butcher series in \cite{ChartierPFI07} and \cite{iserlesBSeriesMethodsCannot2007}. In this setting, trees and forests are replaced by aromatic trees and forests, which allow cycles. Some examples are
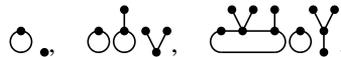
\[
\forestWH , \quad
\forestXH , \quad
\forestYH .
\]
See also \cite{munthe-kaasAromaticButcherSeries2016, bogfjellmo2019}.

Finally, the application of the Butcher series formalism to partitioned and stochastic partitioned differential equations leads to trees and forests with decorations on both vertices and edges. For example,
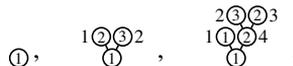
\[
\forestAI \,, \quad
\forestBI \,, \quad
\forestCI \,,
\]
where the label outside a vertex decorates the edge under the vertex. See \cite{BrunedARR19}.

\begin{figure}[h!]
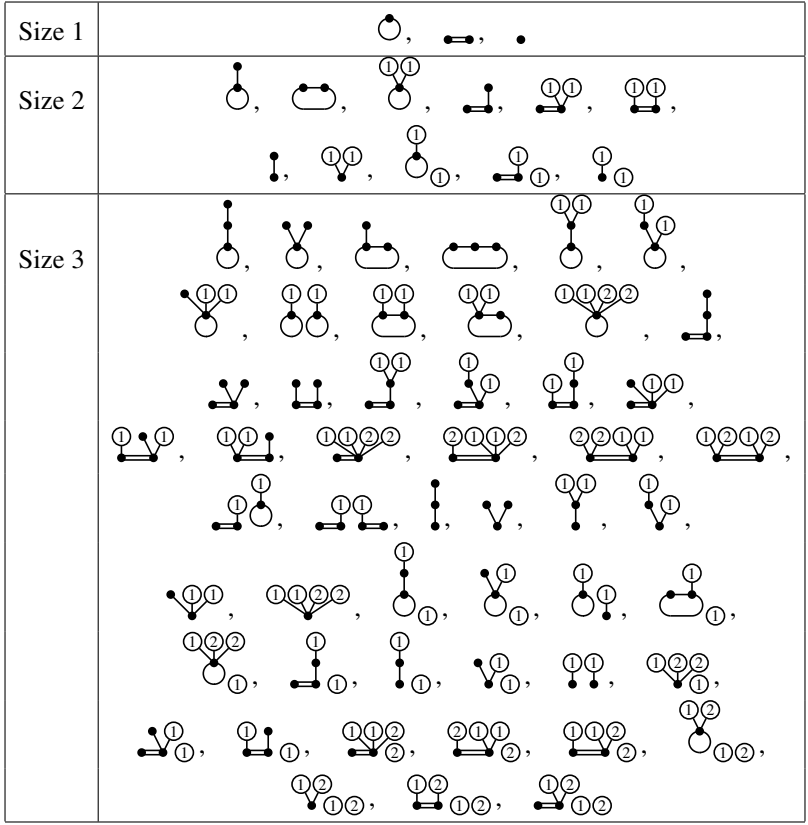

    \centering
    \begin{tabular}{ | c | c |}
        \hline
        Size $1$ & 
        $
            \forestDI \,, \quad
            \forestEI \,, \quad
            \forestFI
        $ \\
        \hline
        Size $2$ &
        $
            \forestGI \,, \quad
            \forestHI  \,, \quad
            \forestII \,, \quad
            \forestJI  \,, \quad
            \forestKI  \,, \quad
            \forestLI  \,,
        $ \\
        &
        $
            \forestMI  \,, \quad
            \forestNI  \,, \quad
            \forestOI  \,, \quad
            \forestPI \,, \quad
            \forestQI 
        $ \\
        \hline
        Size $3$ &
        $
            \forestRI \,, \quad
            \forestSI \,, \quad
            \forestTI \,, \quad
            \forestUI \,, \quad
            \forestVI \,, \quad
           \forestWI \,,
        $ \\
        &
        $
            \forestXI \,, \quad
            \forestYI \,, \quad
            \forestAJ \,, \quad
            \forestBJ \,, \quad
            \forestCJ \,, \quad
            \forestDJ \,,
        $ \\
        &
        $
            \forestEJ \,, \quad
            \forestFJ \,, \quad
            \forestGJ \,, \quad
            \forestHJ \,, \quad
            \forestIJ \,, \quad
            \forestJJ \,,
            
        $ \\
        &
        $
            \forestKJ \,, \quad
            \forestLJ \,, \quad
            \forestMJ \,, \quad
            \forestNJ \,, \quad
            \forestOJ \,, \quad
            \forestPJ \,,
        $ \\
        &
        $
            \forestQJ \,, \quad
            \forestRJ \,, \quad
            \forestSJ \,, \quad
            \forestTJ \,, \quad
            \forestUJ \,, \quad
            \forestVJ \,,
        $ \\
        &
        $
            \forestWJ \,, \quad
            \forestXJ \,, \quad
            \forestYJ \,, \quad
            \forestAK \,, \quad
            \forestBK \,, \quad
            \forestCK \,,
            
        $ \\
        &
        $
            \forestDK \,, \quad
            \forestEK \,, \quad
            \forestFK \,,  \quad
            \forestGK \,, \quad
            \forestHK \,, \quad
            \forestIK \,,
        $ \\
        &
        $
            \forestJK \,, \quad
            \forestKK \,, \quad
            \forestLK \,, \quad
            \forestMK \,, \quad
            \forestNK \,, \quad
            \forestOK \,, 
        $ \\
        &
        $
            \forestPK \,, \quad
            \forestQK \,, \quad
            \forestRK
        $ \\
        \hline
    \end{tabular}
    \caption{All connected exotic aromatic forests with stolons up to size $3$.}
    \label{table:eaforests}
\end{figure}

These extensions can also be combined, leading to highly complex combinatorial objects. As an illustration, \cite{laurentOrderConditionsSampling2022} study exotic aromatic trees and forests with stolons (two horizontal edges connecting two roots). Even listing all such forests up to a small size quickly becomes infeasible by hand. For example, the complete list of all connected exotic aromatic forests with stolons of size up to \(3\) is already nontrivial and is found in Table \ref{table:eaforests}. 

In~\cite{bronascoEfficientLangevinSampling2025}, this framework is further generalized by allowing white vertices to carry incoming edges. This modification leads to a dramatic increase in combinatorial complexity and, consequently, in the number of order conditions: already at order $2$, one obtains a system of 93 conditions. Fortunately, an alternative approach was developed that circumvents the need to work explicitly with these conditions.

\renewcommand{\arraystretch}{1.5}

\subsection{Recursive formulas suitable for implementation}

We observe that Definition \ref{def:grafting} of the grafting product on trees, together with its extension to forests given in \eqref{eq:GuinOudom}, is not ideally suited for implementation. The grafting product on trees is formulated in a combinatorial manner, whereas a recursive definition would lead to a clearer structure and a more natural realization in Haskell. Furthermore, the extension to forests is computationally inefficient, as it involves the generation of intermediate terms in the computation of $(\tau \cdot \pi) \graft \eta$ that cancel out and do not contribute to the final result.

To address these issues, we introduce Definition \ref{def:grafting2}, which presents a recursive definition of the grafting product on forests. This formulation is based on recursion via the Grossman–Larson product and avoids the unnecessary generation of intermediate terms, leading to a clearer and more efficient implementation.

\begin{definition}
    \label{def:grafting2}
    Let \emph{grafting} of a forest and an empty forest be defined as,
    \[ \mathbf{1} \graft \mathbf{1} = \mathbf{1} \,, \quad \pi \graft \mathbf{1} = 0 \,, \quad \mathbf{1} \graft \pi = \pi \,, \quad \text{for } \pi \in F , \pi \neq \one \,. \]
    Grafting of a forest on a tree is defined as,
    \[ \pi \graft B^+_r(\eta) = B^+_r \big(\pi \gl \eta) \,, \quad \text{for } \pi, \eta \in F \,. \]
    Grafting of a forest on a forest is defined as,
    \[ \pi \graft (\tau \cdot \eta) = \sum_{(\pi)} (\pi_{(1)} \graft \tau) \cdot (\pi_{(2)} \graft \eta), \quad \text{for } \tau \in T, \pi, \eta \in F \,. \]
    where $\Delta(\pi) = \sum_{(\pi)} \pi_{(1)} \otimes \pi_{(2)}$ denotes the deshuffle coproduct.
\end{definition}

\section{Implementation of graded vector spaces}
\label{sec:vector_space}

As the core objective of this package is to implement and manipulate algebras of graphs, we first define the fundamental structure on which these algebras are built: graded vector spaces. This section provides the necessary tools for the implementation of the grafting product on decorated forests discussed in Section \ref{sec:algebra}. We start with Assumption \ref{ass:vec_space_graded}.

\begin{assumption}\label{ass:vec_space_graded}
    We consider graded vector spaces $V = \bigoplus_{n \in \mathbb{N}} V_n$ (see Definitions \ref{def:vec_space} and \ref{def:graded_vec_space}) where all homogeneous subspaces $V_n$ are finite-dimensional.
\end{assumption}

Assumption \ref{ass:vec_space_graded} allows us to manipulate formal series as long as the manipulations act nicely with respect to the grading. A vector $v \in V$ can be decomposed as
\[ v = \sum_{n \in \N} v_n \,, \quad \text{with } v_n \in V_n \,, \]
where $V_n$ is the vector space of homogeneous elements with grading $n$. Each $v_n$ is a linear combination of basis elements $\{b_i\}_{i=1}^{d_n}$ with $\text{dim}(V_n) = d_n$, that is,
\[ v_n = \sum_{i=1}^{d_n} c_i b_i \,, \quad \text{for some } c_i \in \R \,. \]
This representation of a vector $v \in V$ is implemented in Implementation \ref{impl:vector}.

\begin{implementation}
    \label{impl:vector}
    We define the elements of a vector space below.
    \begin{haskellcode*}{}
data ScalarProduct k b = SP
    { scalar :: k
    , basisElement :: b
    }
data Sum k b = S Integer (ScalarProduct k b) (Sum k b) | Zero
data Vector k b = V (Sum k b) (Vector k b) | Empty
    \end{haskellcode*}
    where
    \begin{enumerate}
        \item \texttt{ScalarProduct k b} is a pair of a scalar \texttt{k} and a basis element \texttt{b},
        \item \texttt{Sum k b} is a recursively defined list (assumed finite) of scalar products with basis elements of the same grading which is stored in the first argument of \texttt{S},
        \item \texttt{Vector k b} is a recursively defined list (possibly infinite) of sums.
    \end{enumerate}
\end{implementation}

The basis, denoted as \texttt{b}, must be an instance of the \texttt{Eq} and \texttt{Graded} type classes. An instance of \texttt{Eq} ensures that we can determine whether two elements of type \texttt{b} are equal, which is a fundamental requirement for most types. To be an instance of the \texttt{Graded} type class, a type \texttt{b} must implement a function \texttt{grading ::\ b -> Integer}. The definition of the \texttt{Graded} type class is given below.

\begin{implementation}
  \texttt{Graded} type class.
  \begin{haskellcode*}{}
class Graded b where
    grading :: b -> Integer
  \end{haskellcode*}
\end{implementation}

Common types like \texttt{Integer} and \texttt{Char} have instances of the \texttt{Graded} class, where each integer or character is assigned a grading of 1. For composite structures such as lists, 2-tuples, and multisets, the grading is defined as the sum of the gradings of their elements.

The scalar field, \texttt{k}, is required to be an instance of both the \texttt{Num} and \texttt{Eq} type classes. An instance of \texttt{Num} enables arithmetic operations on elements of type \texttt{k}, and most numeric types provide this functionality.

\subsection{Constructing vectors}

Let us detail how the \texttt{ScalarProduct k b}, \texttt{Sum k b}, and \texttt{Vector k b} defined in Implementation \ref{impl:vector} are constructed.
A scalar product \texttt{ScalarProduct k b} and a homogeneous sum \texttt{Sum k b} can be constructed using binary operations \texttt{(*\^{})} and \texttt{(+:)} as can be seen in Usage Example \ref{uex:construction_with_binary}.

\begin{codeexample}
    \label{uex:construction_with_binary}
    We use \texttt{:type} to confirm the types of the expressions.
  \begin{haskellcode*}{}
>>> :type 1 *^ 'x'
1 *^ 'x' :: Num k => ScalarProduct k Char
>>> :type 1*^'x' +: 2*^'y' +: Zero
1*^'x' +: 2*^'y' +: Zero
  :: (Eq k, Num k) => Sum k Char
  \end{haskellcode*}
\end{codeexample}

A vector is constructed from data types that have an instance of the \texttt{IsVector} type class, which includes all types listed in Implementation \ref{impl:vector}. The definiton of the \texttt{IsVector} type class is shown below.

\begin{implementation}
    The \texttt{IsVector} type class leverages \texttt{TypeFamilies}, specifically \texttt{VectorScalar} and \texttt{VectorBasis}, which map the type \texttt{v} to its associated scalar and basis types.
  \begin{haskellcode*}{}
class 
    ( Eq v
    , Eq (VectorBasis v)
    , Graded (VectorBasis v)
    , Eq (VectorScalar v)
    , Num (VectorScalar v)
    )
    => IsVector v where
    type VectorScalar v
    type VectorBasis v
    vector :: v -> Vector (VectorScalar v) (VectorBasis v)
  \end{haskellcode*}
  We note that \texttt{IsVector} is also a \emph{constraint synonym} since it includes a list of common constraints in its definition.
\end{implementation}

As an example, the instance of \texttt{IsVector} for the type \texttt{ScalarProduct k b} is presented below.

\begin{implementation}
  The instance for \texttt{ScalarProduct k b} converts a scalar product into a sum and then uses the instance for \texttt{Sum k b} to obtain the corresponding vector. 
  \begin{haskellcode*}{}
instance (Num k, Eq k, Eq b, Graded b)
        => IsVector (ScalarProduct k b) where
    type VectorScalar (ScalarProduct k b) = k
    type VectorBasis (ScalarProduct k b) = b
    vector = vector . (+: Zero)
  \end{haskellcode*}
\end{implementation}

It is recommended to implement an instance of \texttt{IsVector} for each type \texttt{b} used as a basis in \texttt{Vector k b}. The function \texttt{vector} from the \texttt{IsVector} class can be used to construct finite vectors.

\begin{codeexample}
  Construct vectors from a scalar product and a sum:
  \begin{haskellcode*}{}
>>> :type vector (1 *^ 'x')
vector (1 *^ 'x')
  :: (Num k, Eq k) => Vector k Char
>>> :type vector (1*^'x' +: 2*^'y' +: Zero)
vector (1*^'x' +: 2*^'y' +: Zero)
  :: (Num k, Eq k) => Vector k Char
  \end{haskellcode*}
\end{codeexample}

To construct a finite vector from a finite list of scalar products with varying gradings, use
\[ \texttt{vectorFromList ::\ [ScalarProduct k b] -> Vector k b}. \]
For an infinite vector (formal series), constructed from a potentially infinite list of scalar products with non-decreasing gradings, use
\[ \texttt{vectorFromNonDecList : [ScalarProduct k b] -> Vector k b}. \]

\begin{codeexample}\label{uex:constructing_vectors}
    Use of the \texttt{vectorFromList} function to construct a vector from a finite list of scalar products with varying gradings:
    \begin{haskellcode*}{}
>>> :type vectorFromList [3 *^ "x", 2 *^ "xy", 1 *^ "xyz"]
vectorFromList [3 *^ "x", 2 *^ "xy", 1 *^ "xyz"]
  :: (Num k, Eq k) => Vector k String
    \end{haskellcode*}
    and of the \texttt{vectorFromNonDecList} function to construct a vector from a possibly infinite list of scalar products with non-decreasing gradings:
    \begin{haskellcode*}{}
>>> :type vectorFromNonDecList [i *^ replicate i 'x' | i <- [1..]]
vectorFromNonDecList [i *^ replicate i 'x' | i <- [1..]]
  :: Vector Int [Char]
    \end{haskellcode*}
    where \texttt{takeV n} extracts the first $n$ terms of a vector.
\end{codeexample}

\subsection{Displaying vectors}

The package includes the \texttt{display} function which generates a PDF file with \LaTeX{} representation of \texttt{Vector k b}, see Implementation \ref{impl:display}.

\begin{implementation}
    \label{impl:display}
    Implementation of \texttt{display} function
    \begin{haskellcode*}{}
display
    :: ( IsVector v
       , Texifiable (VectorScalar v)
       , Texifiable (VectorBasis v)
       )
    => v
    -> IO ()
display v = printPdf $ " $ " ++ texify (vector v) ++ " $ "
    \end{haskellcode*}
    where \texttt{printPdf} produces a PDF file with the \LaTeX{} code generated by the \texttt{texify} function from the \texttt{Texifiable} class.
\end{implementation}

The function \texttt{display} assumes the basis \texttt{b} in \texttt{Vector k b} has instance of the \texttt{Texifiable} class whose implementation is not detailed here. The \texttt{Texifiable} class provides a method \texttt{texify ::\ a -> String} that converts an element of type \texttt{a} into its corresponding \LaTeX{} representation.

\begin{codeexample}
    \label{uex:displaying_vectors}
    Use of the \texttt{display} function 
    \begin{haskellcode*}{}
>>> display $ vectorFromList
       [3 *^ "x", 2 *^ "xy", 1 *^ "xyz"]
    \end{haskellcode*}
    $ 3x + 2(x \cdot y) + (x \cdot y \cdot z) $
    \begin{haskellcode*}{}
>>> display $ takeV 5 $ vectorFromNonDecList
       [i *^ replicate i 'x' | i <- [1..]]
    \end{haskellcode*}
    $ x + 2(x \cdot x) + 3(x \cdot x \cdot x) + 4(x \cdot x \cdot x \cdot a) + 5(x \cdot x \cdot x \cdot x \cdot x) $

    where \texttt{takeV n} extracts the first $n$ terms of a vector.
\end{codeexample}

\subsection{Linear and bilinear functions}

Defining linear maps on vector spaces is a common operation in linear algebra and it is often done by starting with a map defined on the basis of the vector space which is then extended linearly to the entire vector space.
To extend a map \texttt{f :: b -> v} defined on the basis \texttt{b} of \texttt{Vector k b} to the entire vector space \texttt{Vector k b}, we use the \texttt{linear} function. It is important to note that \texttt{v} must be a type which has an instance of \texttt{IsVector}.
\begin{codeexample}
    Extending a map defined on the basis \texttt{Char} to a linear map on \texttt{Vector k Char}:
    \begin{haskellcode*}{}
>>> f x = case x of { 'x' -> 1 *^ 'a'; 'y' -> 2 *^ 'b'; 'z' -> 3 *^ 'c' }
>>> v = vector $ 1 *^ 'x' +: 2 *^ 'y' +: 3 *^ 'z' +: Zero
>>> display $ linear f v
    \end{haskellcode*}
    $ a + 4b + 9c $ 
\end{codeexample}

Analogously, products over vector spaces can be defined by starting with a map on the basis elements and extending it bilinearly to the entire vector space. This is achieved using the \texttt{bilinear} function.
\begin{codeexample}
  We use bilinear below:
  \begin{haskellcode*}{}
>>> f x y = 1 *^ [x, y]
>>> v1 = vector
        $ 1 *^ 'x' +: 2 *^ 'y' +: 3 *^ 'z' +: Zero
>>> v2 = vector
        $ 5 *^ 'x' +: 7 *^ 'y' +: 11 *^ 'z' +: Zero
>>> display $ bilinear f v1 v2
  \end{haskellcode*}
    $ 5(x \cdot x) + 10(y \cdot x) + 7(x \cdot y) + 15(z \cdot x) + 14(y \cdot y) + 11(x \cdot z) + 21(z \cdot y) + 22(y \cdot z) + 33(z \cdot z) $
\end{codeexample}

\section{Implementation of algebras of graphs}
\label{sec:algebra}

In this section, we endow the graded vector spaces whose implementation is discussed in Section \ref{sec:vector_space} with algebraic structures introduced in Section \ref{sec:butcher}. In particular, we discuss the symmetric algebra in Section \ref{sec:free_algebras} and the two different approaches in the implementation of algebras of graphs:
\begin{enumerate}
  \item Defining a general graph and subsequently restricting the type of graphs we accept by imposing additional structure,
  \item Building the specific graphs of interest directly from the ground up.
\end{enumerate}

Each approach comes with its own set of advantages and drawbacks. The first approach is more flexible and can accommodate a broader variety of graphs, but it may introduce additional complexity. The second, more specialized approach, is often simpler to implement, although it can require more upfront work to define the specific graphs.

We illustrate the first approach by implementing the grafting algebra of graphs in Section \ref{sec:grafting_of_graphs}. We then focus on the second approach and implement the grafting algebra of trees, forests, as well as the Grossman-Larson algebra in Section \ref{sec:grafting_of_trees}. We finish the section with the implementation of the Connes-Kreimer coalgebra.

\subsection{Symmetric algebra}
\label{sec:free_algebras}

When the basis \texttt{b} of a vector space \texttt{Vector k b} forms a monoid, we can define a product on the vector space \texttt{Vector k b} by extending the product of the monoid to the entire space bilinearly. This turns \texttt{Vector k b} into an algebra with a \texttt{Num} instance which allows us to use familiar arithmetic notation for addition and the product.
The symmetric algebra introduced in Section \ref{sec:butcher:algebra} is implemented as \texttt{Vector k (MultiSet b)}. 

\begin{codeexample}
    We use the $+$ and $*$ operations below for the symmetric algebra \texttt{Vector k (MultiSet b)}.
    \begin{haskellcode*}{}
>>> [x,y,z,a,b,c] = ["x", "y", "z", "a", "b", "c"]
>>> v1 = vector $ 1 *^ x +: 2 *^ y +: 3 *^ z +: Zero
>>> v2 = vector $ 5 *^ a +: 7 *^ b +: 11 *^ c +: Zero
>>> display $ v1 + v2
    \end{haskellcode*}
    $ x + 2y + 3z + 5a + 7b + 11c $ 
    \begin{haskellcode*}{}
>>> display $ v1 * v2
    \end{haskellcode*}
    $ 5(a \cdot x) + 10(a \cdot y) + 7(b \cdot x) + 15(a \cdot z) + 14(b \cdot y) + 11(c \cdot x) + 21(b \cdot z) + 22(c \cdot y) + 33(c \cdot z) $
\end{codeexample}

We define the deshuffle coproduct by following the Definition \ref{def:deshuffle}.

\begin{implementation}
  Implementation of the deshuffle coproduct.
    \begin{haskellcode*}{}
deshuffle
    :: MultiSet b -> Vector Integer (MultiSet b, MultiSet b)
deshuffle =
    product . map (\b -> vector (empty, b') + vector (b', empty))
  where
    b' = singleton b
    \end{haskellcode*}
\end{implementation}

The algebra and coalgebra structure of \texttt{Vector k (MultiSet b)} form the symmetric algebra of the vector space \texttt{Vector k b}, see Section \ref{sec:butcher:symmetric_algebra}.

\subsection{Grafting of graphs}
\label{sec:grafting_of_graphs}

In this section, we discuss the definition and implementation of graphs, along with the grafting operation. We provide an example of implementing an algebra of graphs, starting with a general graph and imposing only the minimal required structure to define the product of interest, which is the grafting product in our case.

\begin{definition}
  \label{def:graph}
  A \emph{graph} $g$ is a set of vertices $V(g)$ together with a set of directed edges $E(g)$. An edge $e$ is defined by its source $s(e) \in V(g)$ and target $t(e) \in V(g)$. A \emph{rooted} graph has a marked vertex called a root.
\end{definition}

We denote the set of graphs by $G$ and the set of rooted graphs by $G_R$.
The implemention of Definition \ref{def:graph} can be found in Implementation \ref{impl:graph}.
\begin{implementation}
    \label{impl:graph}
    A graph is any type \texttt{g} that is an instance of the \texttt{Graph} type class. The \texttt{Graph} type class provides functions for constructing graphs, adding an edge to a graph, and adding graphs together by taking the unions of the sets of vertices and edges.
    \begin{haskellcode*}{}
class Graph g where
    type Vertex g

    singleton :: Vertex g -> g
    edges :: g -> MultiSet (Vertex g, Vertex g)
    vertices :: g -> MultiSet (Vertex g)
    addEdge :: (Vertex g, Vertex g) -> g -> g
    addGraph :: (Graph g0, Vertex g ~ Vertex g0) => g0 -> g -> g

class (Graph g) => RootedGraph g where
    root :: g -> Vertex g
    \end{haskellcode*}
\end{implementation}

A straightforward implementation of a graph which uses integers as vertices is  presented in Implementation \ref{impl:integer_graph}.
\begin{implementation}
  \label{impl:integer_graph}
  A simple implementation of a graph with integers as vertices. Finally, we define the \texttt{Rooted} type and its \texttt{RootedGraph} instance.
    \begin{haskellcode*}{}
data IntegerGraph
    = IG (MultiSet Integer) (MultiSet (Integer, Integer))

instance Graph IntegerGraph where
    type Edge IntegerGraph = (Integer, Integer)

    singleton v = IG (singleton v) empty
    edges (IG _ es) = es
    vertices (IG vs _) = vs
    addGraph g (IG vs es) =
        IG (vertices g `union` vs) (edges g `union` es)
    addEdge e (IG vs es) =
        IG vs (e `insert` es)

data Rooted g = R (Vertex g) g

instance (Graph g) => RootedGraph (Rooted g) where
    root (R r _) = r
  \end{haskellcode*}
\end{implementation}

We define a helper function \texttt{integerGraph} used to construct an \texttt{IntegerGraph} from a list of vertices and edges. We also define the helper function \texttt{rooted} which constructs a rooted graph \texttt{Rooted g} from a graph \texttt{g} by choosing a vertex as a root. We do no discuss the implementations of these helper functions for brevity.

\begin{codeexample}
  \label{uex:graph1}
    An example of a graph:
    \begin{haskellcode*}{}
>>> g = integerGraph [1,2,3] [(1,1),(2,1),(2,3)]
>>> g
IntegerGraph(V=[1,2,3], E=[(1,1),(2,1),(2,3)])
>>> rg = rooted g 1
>>> rg
RootedIntegerGraph(V=[1,2,3], E=[(1,1),(2,1),(2,3)], R=1)
    \end{haskellcode*}

  We note that since we don't assume any structure on the graph, we can't display it in a more visually appealing way.
\end{codeexample}

We are now ready to define and implement a generalization of the grafting product. We rely on Definition \ref{def:grafting} and define the grafting of a rooted graph $g_r \in G_R$ onto a graph $g \in G$ as a sum of all ways to attach the root of $g_r$ to a vertex of $g$.
For example, 
\[ \forestSK \graft \forestTK = \forestUK + \forestVK + \forestWK \,. \]

The implementation of the grafting product as defined in Definition \ref{def:grafting} is presented below.

\begin{implementation}
  Implementation of the grafting product.
  \begin{haskellcode*}{}
graftGraph
    :: ( Eq g2, Graded g2, Graph g2
       , RootedGraph g1, Vertex g1 ~ Vertex g2
       )
    => g1 -> g2 -> Vector Integer g2
graftGraph rg1 g2 =
    vectorFromList $
        map ((1 *^) . graftGraphTo rg1 g2) $
            toList $
                vertices g2

graftGraphTo
    :: ( RootedGraph g1
       , Graph g2, Vertex g1 ~ Vertex g2
       )
    => g1 -> g2 -> Vertex g2 -> g2
graftGraphTo rg1 g2 v
    = addGraph rg1 $ addEdge (root rg1, v) g2
  \end{haskellcode*}
\end{implementation}

We use the \texttt{bilinear} function to extend \texttt{graftGraph} from the basis sets to the corresponding vector spaces to obtain the grafting product. The map \texttt{vectorFromList} constructs a vector as discussed in Section \ref{sec:vector_space}.

\begin{codeexample}
    \label{uex:grafting_of_graphs}
    Example of the grafting product of graphs.
    \begin{haskellcode*}{}
>>> rg1 = rooted (integerGraph [1] []) 1
>>> rg2 = rooted (integerGraph [2] []) 2
>>> g1 = integerGraph [3,4] [(4,3)]
>>> g2 = integerGraph [5] [(5,5)]
>>> bilinear graftGraph
      (vectorFromList [1 *^ rg1, 2 *^ rg2])
      (vectorFromList [3 *^ g1, 4 *^ g2])
(4 *^ IntegerGraph(V=[1,5], E=[(1,5),(5,5)])
      + 8 *^ IntegerGraph(V=[2,5], E=[(2,5),(5,5)]))_2
      + (3 *^ IntegerGraph(V=[1,3,4], E=[(1,3),(4,3)])
      + 3 *^ IntegerGraph(V=[1,3,4], E=[(1,4),(4,3)])
      + 6 *^ IntegerGraph(V=[2,3,4], E=[(2,3),(4,3)])
      + 6 *^ IntegerGraph(V=[2,3,4], E=[(2,4),(4,3)]))_3
    \end{haskellcode*}
    We note that \texttt{rg1} and \texttt{rg2} are rooted graphs while \texttt{g1} and \texttt{g2} are non-rooted.
\end{codeexample}

We recall that the technique presented in this subsection defines a general graph before imposing the structure necessary to define the grafting product. While this approach is flexible and can accommodate a wide variety of graphs, it may also introduce additional complexity and reduce readability.

\subsection{Grafting of forests}
\label{sec:grafting_of_trees}

In this section, we take an alternative approach by constructing the graphs we are interested in from the ground up. This approach allows us to leverage their properties to implement efficient formulas for operations and ensures that the structure we aim to preserve is maintained through these operations which is guaranteed by Haskell's strong and flexible type system. However, this approach requires a more detailed understanding of the graph's structure compared to the one discussed in Section \ref{sec:grafting_of_graphs}.

We will explore the definition and implementation of decorated forests, along with grafting, Grossman-Larson products, and Connes-Kreimer coproduct. We start by implementing trees following Definition \ref{def:tree}. A tree, in this case, is any type \texttt{t} which has an instance of \texttt{IsTree}.

\begin{implementation}
   The \texttt{IsDecorated} type class is used to define the decoration of the tree, while the \texttt{IsTree} type class is used to define the root, children, and construction of the tree.
    \begin{haskellcode*}{}
class IsDecorated a where
    type Decoration a

class (IsDecorated t) => IsTree t where
    root :: t -> Decoration t
    branches :: t -> MultiSet t
    buildTree :: Decoration t -> MultiSet t -> t
    \end{haskellcode*}
\end{implementation}

An implementation of non-planar trees \texttt{Tree d} is presented in Implementation \ref{impl:nonplanar_forests} with forests being represented as multisets of non-planar trees \texttt{MultiSet (Tree d)}.

\begin{implementation}
  \label{impl:nonplanar_forests}
  We define the \texttt{Tree} data type which is an instance of the \texttt{IsDecorated} and \texttt{IsTree} type classes. The \texttt{Tree} data type represents a tree with a root and a multiset of branches.

  \begin{haskellcode*}{}
data Tree d = T
    { nonplanarRoot :: d
    , nonplanarBranches :: MultiSet (Tree d)
    }

instance IsDecorated (Tree d) where
    type Decoration (Tree d) = d

instance IsTree (Tree d) where
    root = nonplanarRoot
    branches = toList . nonplanarChildren

    buildTree r = T r . fromList
    \end{haskellcode*}
\end{implementation}

To facilitate the construction of forests from a textual representation, we provide the \texttt{fromBrackets} function, which is part of the \texttt{HasBracketNotation} type class. This function takes a single string representing a forest in bracket notation and converts it into the corresponding Haskell forest type. For example, given a forest encoded as a string in bracket notation, the expression
\begin{haskellcode*}{}
fromBrackets "1[2,3],4[5[6]],7" :: MultiSet (Tree Integer)
\end{haskellcode*}
produces a forest represented as a multiset of \texttt{Tree Integer}.

\subsubsection{Grafting and Grossman-Larson products}

Implementation \ref{impl:grafting} provides the implementation of the grafting and Grossman-Larson products of forests using the Definition \ref{def:grafting2}.

\begin{implementation}
  \label{impl:grafting}
  To facilitate future extension of the algebras, we define the \texttt{CanGraft} and \texttt{CanGrossmanLarson} type classes for graphs that support grafting and Grossman-Larson products.
    \begin{haskellcode*}{}
class (IsVector a) => CanGraft a where
    graft
        :: a -> a -> Vector (VectorScalar a) (VectorBasis a)

class (IsVector a) => CanGrossmanLarson a where
    grossmanLarson
        :: a -> a -> Vector (VectorScalar a) (VectorBasis a)

instance
    ( IsTree t
    , IsVector t
    ) => CanGraft (MultiSet t)
    where
    graft empty empty = vector empty
    graft _ empty = vector Zero
    graft empty f2 = vector f2
    graft f1 f2 = case (toList f2) of
        [t]    -> linear (singleton . buildTree (root t))
                    $ grossmanLarson f $ branches t
        t : ts -> linear perTerm $ deshuffle f1
      where
        perTerm (f11, f12)
            = graft f11 (singleton t) * graft f12 ts

instance 
    ( IsTree t
    , IsVector t
    )
    => CanGrossmanLarson (MultiSet t)
    where
    grossmanLarson f1 f2 = linear perTerm $ deshuffle f1
      where
        perTerm (f11, f12)
            = vector f11 * graft f12 f2

    \end{haskellcode*}
\end{implementation}
 
An example of the use of forests can be found in Usage Example \ref{uex:nonplanar_forests} in the \emph{Getting Started} section of the paper as well as in Usage Example \ref{uex:grafting}.

\begin{codeexample}
  \label{uex:grafting}
    We define the trees $\forestXK$ and $\forestYK$ and compute their grafting product.
    \begin{haskellcode*}{}
>>> f1 = fromBrackets "1[2],3" :: MultiSet (Tree Integer)
>>> f2 = fromBrackets "4[5,6]" :: MultiSet (Tree Integer)
>>> display $ graft f1 f2
  \end{haskellcode*}
  $\forestAL + \forestBL + \forestCL + \forestDL + \forestEL + \forestFL + \forestGL + \forestHL + \forestIL$
\end{codeexample}

\subsubsection{Connes-Kreimer coproduct}
\label{sec:CK}

We implement Connes-Kreimer coproduct by following Definition \ref{def:ck}.

\begin{implementation}
  \label{impl:ck}
  To facilitate future extension of the Connes-Kreimer coalgebra, we define the \texttt{CanConnesKreimer} type class for graphs that support Connes-Kreimer coproduct.
    \begin{haskellcode*}{}
class (IsVector a) => CanConnesKreimer a where
    connesKreimer
        :: a -> Vector (VectorScalar a)
                    (VectorBasis a, VectorBasis a)

instance
    ( Ord t
    , IsVector t
    , IsTree t
    ) => CanConnesKreimer (MS.MultiSet t) where
    connesKreimer ms = case (toList ms) of
        []  -> vector (empty, empty)
        [t] -> vector (singleton t, empty)
                + (linear perTerm
                    $ connesKreimer $ fromList $ branches t)
          where
            perTerm (f1, f2) = (f1, singleton
                                        $ buildTree r $ toList f2)
            r = root t
        f  -> product
                $ map (connesKreimer . singleton) f

    \end{haskellcode*}
\end{implementation}

\section{Conclusion}
\label{sec:conclusion}

In this work, we have presented \texttt{Arboretum.hs} as a flexible and type-safe framework for symbolic computations with algebras of trees and forests built on top of graded vector spaces, designed to closely mirror their mathematical structure while remaining accessible and extensible. In contrast to existing software such as \texttt{BSeries.jl}, which focuses on efficient implementations of classical Butcher series built from classical rooted trees and is primarily tailored to established applications in numerical analysis, \texttt{Arboretum.hs} adopts a broader and more exploratory perspective, accommodating more general combinatorial structures and algebraic operations beyond the classical setting. Moreover, while \texttt{BSeries.jl} benefits from the performance and ecosystem of Julia, it does not leverage a static type system to encode algebraic invariants, leaving certain classes of errors to be detected only at runtime; by contrast, Haskell’s strong and expressive type system allows \texttt{Arboretum.hs} to reflect algebraic structures directly in the code and enforce correctness properties at compile time. As a result, the package provides a robust environment not only for computation but also for experimentation and verification, making it a useful tool for researchers working at the interface of algebra, combinatorics, and applications.

For the sake of exposition, we have focused on the implementation of the pre-Lie grafting algebra of decorated forests within the \texttt{Arboretum.hs} package. However, the design principles and modular structure of the package allow for straightforward extensions to other algebraic structures and operations on graphs. In particular, the package includes the following implementations that were not discussed in detail in the main text:
\begin{enumerate}
    \item Shuffle-deconcatenation Hopf algebra,
    \item Planar forests and trees together with grafting, Grossman-Larson products, Connes-Kreimer coproduct, and other operations for which recursive formulas suitable for implementation were derived in \cite{munthe-kaas2018},
    \item Aromatic forests and trees together with the grafting, Grossman-Larson, substitution products, as well as the divergence operation. See \cite{floystadUniversalPreLieRinehartAlgebras2021,bronascoHAS24}. 
    \item Syntactic trees which correspond to the mathematical concept of operads or computer science concept of abstract syntax trees.
\end{enumerate}

Planned features include:
\begin{enumerate}
    \item Implementation of exotic forests and the corresponding products and operations, see \cite{laurentOrderConditionsSampling2022, BronascoEBS25},
    \item Implementation the algorithm for the generation of order conditions for invariant measure sampling of overdamped Langevin dynamics, see \cite{laurentOrderConditionsSampling2022,BronascoEBS25}.
    \item Integration of the package \texttt{Operads}\footnote{\url{https://hackage.haskell.org/package/Operads}} which computes Gr\"{o}bner basis for operads (see \cite{DotsenkoGBO10}) into the \texttt{Arboretum.hs} package.
\end{enumerate}

We believe this package will serve as a useful tool for both practitioners and theoretical researchers working at the interface of algebra, combinatorics, and applications.

\section*{Acknowledgements}

EB acknowledges the support of the Knut and Alice Wallenberg Foundation (grant number KAW 2023.0433).
EB and GV acknowledge the support of the Swiss National Science Foundation, projects No 200020\_214819, No. 200020\_192129 and No. 10009199.


\bibliography{references,zotero}

\label{lastpage01}

\end{document}

%% file: planarforest_tikz.tex
\newcommand{\forestA}{
\tikz[planar forest] {

\node [lc] at (0.0, 0.0) { 1 } 
child {node [lc] at (0.0, 1.0) { 2 }  
}
;
}}

\newcommand{\forestB}{
\tikz[planar forest] {

\node [lc] at (0.0, 0.0) { 1 } 
child {node [lc] at (0.0, 1.0) { 2 }  
}
;
}}

\newcommand{\forestC}{
\tikz[planar forest] {

\node [lc] at (0.0, 0.0) { 1 } 
;
}}

\newcommand{\forestD}{
\tikz[planar forest] {

\node [lc] at (0.0, 0.0) { 1 } 
child {node [lc] at (-0.5, 1.0) { 2 }  
}
child {node [lc] at (0.5, 1.0) { 2 }  
}
;
}}

\newcommand{\forestE}{
\tikz[planar forest] {

\node [lc] at (0.0, 0.0) { 1 } 
;
}}

\newcommand{\forestF}{
\tikz[planar forest] {

\node [lc] at (0.0, 0.0) { 1 } 
child {node [lc] at (-0.5, 1.0) { 2 }  
child {node [lc] at (0.0, 1.0) { 1 }  
child {node [lc] at (0.0, 1.0) { 2 }  
}
}
}
child {node [lc] at (0.5, 1.0) { 2 }  
child {node [lc] at (0.0, 1.0) { 1 }  
child {node [lc] at (0.0, 1.0) { 2 }  
}
}
}
;
}}

\newcommand{\forestG}{
\tikz[planar forest] {

\node [lc] at (0.0, 0.0) { 1 } 
;
}}

\newcommand{\forestH}{
\tikz[planar forest] {

\node [lc] at (0.0, 0.0) { 1 } 
child {node [lc] at (-0.5, 1.0) { 2 }  
}
child {node [lc] at (0.5, 1.0) { 2 }  
child {node [lc] at (-0.5, 1.0) { 1 }  
child {node [lc] at (0.0, 1.0) { 2 }  
}
}
child {node [lc] at (0.5, 1.0) { 1 }  
child {node [lc] at (0.0, 1.0) { 2 }  
}
}
}
;
}}

\newcommand{\forestI}{
\tikz[planar forest] {

\node [lc] at (0.0, 0.0) { 1 } 
;
}}

\newcommand{\forestJ}{
\tikz[planar forest] {

\node [lc] at (0.0, 0.0) { 1 } 
child {node [lc] at (-1.0, 1.0) { 1 }  
child {node [lc] at (0.0, 1.0) { 2 }  
}
}
child {node [lc] at (0.0, 1.0) { 2 }  
}
child {node [lc] at (1.0, 1.0) { 2 }  
child {node [lc] at (0.0, 1.0) { 1 }  
child {node [lc] at (0.0, 1.0) { 2 }  
}
}
}
;
}}

\newcommand{\forestK}{
\tikz[planar forest] {

\node [lc] at (0.0, 0.0) { 1 } 
;
}}

\newcommand{\forestL}{
\tikz[planar forest] {

\node [lc] at (0.0, 0.0) { 1 } 
child {node [lc] at (-1.5, 1.0) { 1 }  
child {node [lc] at (0.0, 1.0) { 2 }  
}
}
child {node [lc] at (-0.5, 1.0) { 1 }  
child {node [lc] at (0.0, 1.0) { 2 }  
}
}
child {node [lc] at (0.5, 1.0) { 2 }  
}
child {node [lc] at (1.5, 1.0) { 2 }  
}
;
}}

\newcommand{\forestM}{
\tikz[planar forest] {

\node [lc] at (0.0, 0.0) { 1 } 
child {node [lc] at (0.0, 1.0) { 1 }  
child {node [lc] at (0.0, 1.0) { 2 }  
}
}
;
}}

\newcommand{\forestN}{
\tikz[planar forest] {

\node [lc] at (0.0, 0.0) { 1 } 
child {node [lc] at (-0.5, 1.0) { 2 }  
}
child {node [lc] at (0.5, 1.0) { 2 }  
child {node [lc] at (0.0, 1.0) { 1 }  
child {node [lc] at (0.0, 1.0) { 2 }  
}
}
}
;
}}

\newcommand{\forestO}{
\tikz[planar forest] {

\node [lc] at (0.0, 0.0) { 1 } 
child {node [lc] at (0.0, 1.0) { 1 }  
child {node [lc] at (0.0, 1.0) { 2 }  
}
}
;
}}

\newcommand{\forestP}{
\tikz[planar forest] {

\node [lc] at (0.0, 0.0) { 1 } 
child {node [lc] at (-1.0, 1.0) { 1 }  
child {node [lc] at (0.0, 1.0) { 2 }  
}
}
child {node [lc] at (0.0, 1.0) { 2 }  
}
child {node [lc] at (1.0, 1.0) { 2 }  
}
;
}}

\newcommand{\forestQ}{
\tikz[planar forest] {

\node [lc] at (0.0, 0.0) { 1 } 
child {node [lc] at (-0.5, 1.0) { 1 }  
child {node [lc] at (0.0, 1.0) { 2 }  
}
}
child {node [lc] at (0.5, 1.0) { 1 }  
child {node [lc] at (0.0, 1.0) { 2 }  
}
}
;
}}

\newcommand{\forestR}{
\tikz[planar forest] {

\node [lc] at (0.0, 0.0) { 1 } 
child {node [lc] at (-0.5, 1.0) { 2 }  
}
child {node [lc] at (0.5, 1.0) { 2 }  
}
;
}}

\newcommand{\forestS}{
\tikz[planar forest] {

\node [b] at (0.0, 0.0) {  } 
;
}}

\newcommand{\forestT}{
\tikz[planar forest] {

\node [b] at (0.0, 0.0) {  } 
child {node [b] at (0.0, 1.0) {  }  
}
;
}}

\newcommand{\forestU}{
\tikz[planar forest] {

\node [b] at (0.0, 0.0) {  } 
child {node [b] at (-0.5, 1.0) {  }  
}
child {node [b] at (0.5, 1.0) {  }  
}
;
}}

\newcommand{\forestV}{
\tikz[planar forest] {

\node [b] at (0.0, 0.0) {  } 
child {node [b] at (0.0, 1.0) {  }  
child {node [b] at (0.0, 1.0) {  }  
}
}
;
}}

\newcommand{\forestW}{
\tikz[planar forest] {

\node [b] at (0.0, 0.0) {  } 
child {node [b] at (-1.0, 1.0) {  }  
}
child {node [b] at (0.0, 1.0) {  }  
}
child {node [b] at (1.0, 1.0) {  }  
}
;
}}

\newcommand{\forestX}{
\tikz[planar forest] {

\node [b] at (0.0, 0.0) {  } 
child {node [b] at (-0.5, 1.0) {  }  
child {node [b] at (0.0, 1.0) {  }  
}
}
child {node [b] at (0.5, 1.0) {  }  
}
;
}}

\newcommand{\forestY}{
\tikz[planar forest] {

\node [b] at (0.0, 0.0) {  } 
child {node [b] at (0.0, 1.0) {  }  
child {node [b] at (-0.5, 1.0) {  }  
}
child {node [b] at (0.5, 1.0) {  }  
}
}
;
}}

\newcommand{\forestAB}{
\tikz[planar forest] {

\node [b] at (0.0, 0.0) {  } 
child {node [b] at (0.0, 1.0) {  }  
child {node [b] at (0.0, 1.0) {  }  
child {node [b] at (0.0, 1.0) {  }  
}
}
}
;
}}

\newcommand{\forestBB}{
\tikz[planar forest] {

\node [b] at (0.0, 0.0) {  } 
child {node [b] at (-0.5, 1.0) {  }  
child {node [b] at (0.0, 1.0) {  }  
}
}
child {node [b] at (0.5, 1.0) {  }  
}
;
}}

\newcommand{\forestCB}{
\tikz[planar forest] {

\node [b] at (0.0, 0.0) {  } 
child {node [b] at (-0.5, 1.0) {  }  
}
child {node [b] at (0.5, 1.0) {  }  
child {node [b] at (0.0, 1.0) {  }  
}
}
;
}}

\newcommand{\forestDB}{
\tikz[planar forest] {

\node [b] at (0.0, 0.0) {  } 
child {node [b] at (0.0, 1.0) {  }  
}
;
}}

\newcommand{\forestEB}{
\tikz[planar forest] {

\node [b] at (0.0, 0.0) {  } 
child {node [b] at (-0.5, 1.0) {  }  
}
child {node [b] at (0.5, 1.0) {  }  
}
;
\node [b] at (1.0, 0.0) {  } 
;
\node [b] at (2.0, 0.0) {  } 
;
}}

\newcommand{\forestFB}{
\tikz[planar forest] {

\node [b] at (0.0, 0.0) {  } 
child {node [b] at (-0.5, 1.0) {  }  
}
child {node [b] at (0.5, 1.0) {  }  
child {node [b] at (0.0, 1.0) {  }  
}
}
;
\node [b] at (1.5, 0.0) {  } 
child {node [b] at (0.0, 1.0) {  }  
}
;
\node [b] at (2.5, 0.0) {  } 
;
\node [b] at (3.5, 0.0) {  } 
child {node [b] at (0.0, 1.0) {  }  
child {node [b] at (0.0, 1.0) {  }  
}
}
;
}}

\newcommand{\forestGB}{
\tikz[planar forest] {

\node [b] at (0.0, 0.0) {  } 
child {node [b] at (0.0, 1.0) {  }  
}
;
\node [b] at (1.0, 0.0) {  } 
;
}}

\newcommand{\forestHB}{
\tikz[planar forest] {

\node [b] at (0.0, 0.0) {  } 
;
\node [b] at (1.0, 0.0) {  } 
child {node [b] at (0.0, 1.0) {  }  
}
;
}}

\newcommand{\forestIB}{
\tikz[planar forest] {

\node [b] at (0.0, 0.0) {  } 
;
}}

\newcommand{\forestJB}{
\tikz[planar forest] {

\node [b] at (0.0, 0.0) {  } 
child {node [b] at (0.0, 1.0) {  }  
}
;
}}

\newcommand{\forestKB}{
\tikz[planar forest] {

\node [b] at (0.0, 0.0) {  } 
child {node [b] at (0.0, 1.0) {  }  
child {node [b] at (0.0, 1.0) {  }  
}
}
;
}}

\newcommand{\forestLB}{
\tikz[planar forest] {

\node [b] at (0.0, 0.0) {  } 
child {node [b] at (0.0, 1.0) {  }  
}
;
\node [b] at (1.0, 0.0) {  } 
;
\node [b] at (2.0, 0.0) {  } 
;
}}

\newcommand{\forestMB}{
\tikz[planar forest] {

\node [b] at (0.0, 0.0) {  } 
child {node [b] at (-1.0, 1.0) {  }  
child {node [b] at (0.0, 1.0) {  }  
}
}
child {node [b] at (0.0, 1.0) {  }  
}
child {node [b] at (1.0, 1.0) {  }  
}
;
}}

\newcommand{\forestNB}{
\tikz[planar forest] {

\node [b] at (0.0, 0.0) {  } 
;
}}

\newcommand{\forestOB}{
\tikz[planar forest] {

\node [b] at (0.0, 0.0) {  } 
child {node [b] at (0.0, 1.0) {  }  
}
;
}}

\newcommand{\forestPB}{
\tikz[planar forest] {

\node [b] at (0.0, 0.0) {  } 
child {node [b] at (-0.5, 1.0) {  }  
}
child {node [b] at (0.5, 1.0) {  }  
child {node [b] at (0.0, 1.0) {  }  
}
}
;
}}

\newcommand{\forestQB}{
\tikz[planar forest] {

\node [b] at (0.0, 0.0) {  } 
;
}}

\newcommand{\forestRB}{
\tikz[planar forest] {

\node [b] at (0.0, 0.0) {  } 
child {node [b] at (0.0, 1.0) {  }  
}
;
}}

\newcommand{\forestSB}{
\tikz[planar forest] {

\node [b] at (0.0, 0.0) {  } 
child {node [b] at (-0.5, 1.0) {  }  
}
child {node [b] at (0.5, 1.0) {  }  
}
;
}}

\newcommand{\forestTB}{
\tikz[planar forest] {

\node [b] at (0.0, 0.0) {  } 
child {node [b] at (0.0, 1.0) {  }  
child {node [b] at (0.0, 1.0) {  }  
}
}
;
}}

\newcommand{\forestUB}{
\tikz[planar forest] {

\node [b] at (0.0, 0.0) {  } 
child {node [b] at (-1.0, 1.0) {  }  
}
child {node [b] at (0.0, 1.0) {  }  
}
child {node [b] at (1.0, 1.0) {  }  
}
;
}}

\newcommand{\forestVB}{
\tikz[planar forest] {

\node [b] at (0.0, 0.0) {  } 
child {node [b] at (-0.5, 1.0) {  }  
child {node [b] at (0.0, 1.0) {  }  
}
}
child {node [b] at (0.5, 1.0) {  }  
}
;
}}

\newcommand{\forestWB}{
\tikz[planar forest] {

\node [b] at (0.0, 0.0) {  } 
child {node [b] at (0.0, 1.0) {  }  
child {node [b] at (-0.5, 1.0) {  }  
}
child {node [b] at (0.5, 1.0) {  }  
}
}
;
}}

\newcommand{\forestXB}{
\tikz[planar forest] {

\node [b] at (0.0, 0.0) {  } 
child {node [b] at (0.0, 1.0) {  }  
child {node [b] at (0.0, 1.0) {  }  
child {node [b] at (0.0, 1.0) {  }  
}
}
}
;
}}

\newcommand{\forestYB}{
\tikz[planar forest] {

\node [b] at (0.0, 0.0) {  } 
;
}}

\newcommand{\forestAC}{
\tikz[planar forest] {

\node [b] at (0.0, 0.0) {  } 
child {node [b] at (0.0, 1.0) {  }  
}
;
}}

\newcommand{\forestBC}{
\tikz[planar forest] {

\node [b] at (0.0, 0.0) {  } 
child {node [b] at (-0.5, 1.0) {  }  
}
child {node [b] at (0.5, 1.0) {  }  
}
;
}}

\newcommand{\forestCC}{
\tikz[planar forest] {

\node [b] at (0.0, 0.0) {  } 
child {node [b] at (0.0, 1.0) {  }  
child {node [b] at (0.0, 1.0) {  }  
}
}
;
}}

\newcommand{\forestDC}{
\tikz[planar forest] {

\node [b] at (0.0, 0.0) {  } 
child {node [b] at (-1.0, 1.0) {  }  
}
child {node [b] at (0.0, 1.0) {  }  
}
child {node [b] at (1.0, 1.0) {  }  
}
;
}}

\newcommand{\forestEC}{
\tikz[planar forest] {

\node [b] at (0.0, 0.0) {  } 
child {node [b] at (-0.5, 1.0) {  }  
child {node [b] at (0.0, 1.0) {  }  
}
}
child {node [b] at (0.5, 1.0) {  }  
}
;
}}

\newcommand{\forestFC}{
\tikz[planar forest] {

\node [b] at (0.0, 0.0) {  } 
child {node [b] at (0.0, 1.0) {  }  
child {node [b] at (-0.5, 1.0) {  }  
}
child {node [b] at (0.5, 1.0) {  }  
}
}
;
}}

\newcommand{\forestGC}{
\tikz[planar forest] {

\node [b] at (0.0, 0.0) {  } 
child {node [b] at (0.0, 1.0) {  }  
child {node [b] at (0.0, 1.0) {  }  
child {node [b] at (0.0, 1.0) {  }  
}
}
}
;
}}

\newcommand{\forestHC}{
\tikz[planar forest] {

\node [b] at (0.0, 0.0) {  } 
;
}}

\newcommand{\forestIC}{
\tikz[planar forest] {

\node [b] at (0.0, 0.0) {  } 
child {node [b] at (0.0, 1.0) {  }  
}
;
}}

\newcommand{\forestJC}{
\tikz[planar forest] {

\node [b] at (0.0, 0.0) {  } 
child {node [b] at (-0.5, 1.0) {  }  
}
child {node [b] at (0.5, 1.0) {  }  
}
;
}}

\newcommand{\forestKC}{
\tikz[planar forest] {

\node [b] at (0.0, 0.0) {  } 
child {node [b] at (0.0, 1.0) {  }  
child {node [b] at (0.0, 1.0) {  }  
}
}
;
}}

\newcommand{\forestLC}{
\tikz[planar forest] {

\node [b] at (0.0, 0.0) {  } 
child {node [b] at (0.0, 1.0) {  }  
}
;
}}

\newcommand{\forestMC}{
\tikz[planar forest] {

\node [b] at (0.0, 0.0) {  } 
child {node [b] at (-0.5, 1.0) {  }  
}
child {node [b] at (0.5, 1.0) {  }  
child {node [b] at (0.0, 1.0) {  }  
}
}
;
}}

\newcommand{\forestNC}{
\tikz[planar forest] {

\node [b] at (0.0, 0.0) {  } 
child {node [b] at (-1.0, 1.0) {  }  
child {node [b] at (0.0, 1.0) {  }  
}
}
child {node [b] at (0.0, 1.0) {  }  
}
child {node [b] at (1.0, 1.0) {  }  
child {node [b] at (0.0, 1.0) {  }  
}
}
;
}}

\newcommand{\forestOC}{
\tikz[planar forest] {

\node [b] at (0.0, 0.0) {  } 
child {node [b] at (-0.5, 1.0) {  }  
child {node [b] at (0.0, 1.0) {  }  
child {node [b] at (0.0, 1.0) {  }  
}
}
}
child {node [b] at (0.5, 1.0) {  }  
child {node [b] at (0.0, 1.0) {  }  
}
}
;
}}

\newcommand{\forestPC}{
\tikz[planar forest] {

\node [b] at (0.0, 0.0) {  } 
child {node [b] at (-0.5, 1.0) {  }  
}
child {node [b] at (0.5, 1.0) {  }  
child {node [b] at (-0.5, 1.0) {  }  
child {node [b] at (0.0, 1.0) {  }  
}
}
child {node [b] at (0.5, 1.0) {  }  
}
}
;
}}

\newcommand{\forestQC}{
\tikz[planar forest] {

\node [b] at (0.0, 0.0) {  } 
child {node [b] at (-0.5, 1.0) {  }  
}
child {node [b] at (0.5, 1.0) {  }  
child {node [b] at (0.0, 1.0) {  }  
child {node [b] at (0.0, 1.0) {  }  
child {node [b] at (0.0, 1.0) {  }  
}
}
}
}
;
}}

\newcommand{\forestRC}{
\tikz[planar forest] {

\node [b] at (0.0, 0.0) {  } 
child {node [b] at (0.0, 1.0) {  }  
}
;
}}

\newcommand{\forestSC}{
\tikz[planar forest] {

\node [b] at (0.0, 0.0) {  } 
child {node [b] at (-0.5, 1.0) {  }  
child {node [b] at (0.0, 1.0) {  }  
}
}
child {node [b] at (0.5, 1.0) {  }  
}
;
}}

\newcommand{\forestTC}{
\tikz[planar forest] {

\node [b] at (0.0, 0.0) {  } 
child {node [b] at (0.0, 1.0) {  }  
}
;
}}

\newcommand{\forestUC}{
\tikz[planar forest] {

\node [b] at (0.0, 0.0) {  } 
child {node [b] at (0.0, 1.0) {  }  
}
;
}}

\newcommand{\forestVC}{
\tikz[planar forest] {

\node [b] at (0.0, 0.0) {  } 
child {node [b] at (0.0, 1.0) {  }  
child {node [b] at (0.0, 1.0) {  }  
child {node [b] at (0.0, 1.0) {  }  
}
}
}
;
}}

\newcommand{\forestWC}{
\tikz[planar forest] {

\node [b] at (0.0, 0.0) {  } 
child {node [b] at (0.0, 1.0) {  }  
child {node [b] at (0.0, 1.0) {  }  
}
}
;
}}

\newcommand{\forestXC}{
\tikz[planar forest] {

\node [b] at (0.0, 0.0) {  } 
child {node [b] at (0.0, 1.0) {  }  
}
;
}}

\newcommand{\forestYC}{
\tikz[planar forest] {

\node [b] at (0.0, 0.0) {  } 
child {node [b] at (0.0, 1.0) {  }  
child {node [b] at (-0.5, 1.0) {  }  
}
child {node [b] at (0.5, 1.0) {  }  
}
}
;
}}

\newcommand{\forestAD}{
\tikz[planar forest] {

\node [b] at (0.0, 0.0) {  } 
child {node [b] at (-0.5, 1.0) {  }  
}
child {node [b] at (0.5, 1.0) {  }  
}
;
}}

\newcommand{\forestBD}{
\tikz[planar forest] {

\node [b] at (0.0, 0.0) {  } 
child {node [b] at (-0.5, 1.0) {  }  
}
child {node [b] at (0.5, 1.0) {  }  
}
;
}}

\newcommand{\forestCD}{
\tikz[planar forest] {

\node [b] at (0.0, 0.0) {  } 
child {node [b] at (0.0, 1.0) {  }  
}
;
}}

\newcommand{\forestDD}{
\tikz[planar forest] {

\node [b] at (0.0, 0.0) {  } 
child {node [b] at (0.0, 1.0) {  }  
child {node [b] at (0.0, 1.0) {  }  
}
}
;
}}

\newcommand{\forestED}{
\tikz[planar forest] {

\node [b] at (0.0, 0.0) {  } 
child {node [b] at (-1.0, 1.0) {  }  
}
child {node [b] at (0.0, 1.0) {  }  
}
child {node [b] at (1.0, 1.0) {  }  
}
;
}}

\newcommand{\forestFD}{
\tikz[planar forest] {

\node [b] at (0.0, 0.0) {  } 
child {node [b] at (-0.5, 1.0) {  }  
}
child {node [b] at (0.5, 1.0) {  }  
}
;
}}

\newcommand{\forestGD}{
\tikz[planar forest] {

\node [b] at (0.0, 0.0) {  } 
child {node [b] at (-0.5, 1.0) {  }  
child {node [b] at (0.0, 1.0) {  }  
}
}
child {node [b] at (0.5, 1.0) {  }  
}
;
}}

\newcommand{\forestHD}{
\tikz[planar forest] {

\node [b] at (0.0, 0.0) {  } 
child {node [b] at (0.0, 1.0) {  }  
child {node [b] at (0.0, 1.0) {  }  
child {node [b] at (0.0, 1.0) {  }  
}
}
}
;
}}

\newcommand{\forestID}{
\tikz[planar forest] {

\node [b] at (0.0, 0.0) {  } 
child {node [b] at (0.0, 1.0) {  }  
child {node [b] at (0.0, 1.0) {  }  
}
}
;
}}

\newcommand{\forestJD}{
\tikz[planar forest] {

\node [b] at (0.0, 0.0) {  } 
;
}}

\newcommand{\forestKD}{
\tikz[planar forest] {

\node [b] at (0.0, 0.0) {  } 
;
}}

\newcommand{\forestLD}{
\tikz[planar forest] {

\node [b] at (0.0, 0.0) {  } 
;
}}

\newcommand{\forestMD}{
\tikz[planar forest] {

\node [b] at (0.0, 0.0) {  } 
;
}}

\newcommand{\forestND}{
\tikz[planar forest] {

\node [b] at (0.0, 0.0) {  } 
;
\node [b] at (1.0, 0.0) {  } 
child {node [b] at (0.0, 1.0) {  }  
}
;
}}

\newcommand{\forestOD}{
\tikz[planar forest] {

\node [b] at (0.0, 0.0) {  } 
;
\node [b] at (1.0, 0.0) {  } 
;
}}

\newcommand{\forestPD}{
\tikz[planar forest] {

\node [b] at (0.0, 0.0) {  } 
child {node [b] at (-0.5, 1.0) {  }  
}
child {node [b] at (0.5, 1.0) {  }  
child {node [b] at (0.0, 1.0) {  }  
}
}
;
\node [b] at (1.0, 0.0) {  } 
;
}}

\newcommand{\forestQD}{
\tikz[planar forest] {

\node [b] at (0.0, 0.0) {  } 
child {node [b] at (0.0, 1.0) {  }  
}
;
\node [b] at (1.0, 0.0) {  } 
child {node [b] at (0.0, 1.0) {  }  
child {node [b] at (0.0, 1.0) {  }  
}
}
;
}}

\newcommand{\forestRD}{
\tikz[planar forest] {

\node [b] at (0.0, 0.0) {  } 
;
\node [b] at (1.0, 0.0) {  } 
child {node [b] at (0.0, 1.0) {  }  
}
;
}}

\newcommand{\forestSD}{
\tikz[planar forest] {

\node [b] at (0.0, 0.0) {  } 
child {node [b] at (0.0, 1.0) {  }  
}
;
}}

\newcommand{\forestTD}{
\tikz[planar forest] {

\node [b] at (0.0, 0.0) {  } 
child {node [b] at (-1.0, 1.0) {  }  
}
child {node [b] at (0.0, 1.0) {  }  
child {node [b] at (0.0, 1.0) {  }  
}
}
child {node [b] at (1.0, 1.0) {  }  
}
;
}}

\newcommand{\forestUD}{
\tikz[planar forest] {

\node [b] at (0.0, 0.0) {  } 
child {node [b] at (-0.5, 1.0) {  }  
child {node [b] at (0.0, 1.0) {  }  
}
}
child {node [b] at (0.5, 1.0) {  }  
child {node [b] at (0.0, 1.0) {  }  
}
}
;
}}

\newcommand{\forestVD}{
\tikz[planar forest] {

\node [b] at (0.0, 0.0) {  } 
child {node [b] at (-0.5, 1.0) {  }  
}
child {node [b] at (0.5, 1.0) {  }  
child {node [b] at (0.0, 1.0) {  }  
child {node [b] at (0.0, 1.0) {  }  
}
}
}
;
}}

\newcommand{\forestWD}{
\tikz[planar forest] {

\node [b] at (0.0, 0.0) {  } 
child {node [b] at (0.0, 1.0) {  }  
child {node [b] at (-0.5, 1.0) {  }  
}
child {node [b] at (0.5, 1.0) {  }  
child {node [b] at (0.0, 1.0) {  }  
}
}
}
;
}}

\newcommand{\forestXD}{
\tikz[planar forest] {

\node [b] at (0.0, 0.0) {  } 
;
}}

\newcommand{\forestYD}{
\tikz[planar forest] {

\node [b] at (0.0, 0.0) {  } 
child {node [b] at (0.0, 1.0) {  }  
}
;
}}

\newcommand{\forestAE}{
\tikz[planar forest] {

\node [b] at (0.0, 0.0) {  } 
;
\node [b] at (1.0, 0.0) {  } 
child {node [b] at (0.0, 1.0) {  }  
}
;
}}

\newcommand{\forestBE}{
\tikz[planar forest] {

\node [b] at (0.0, 0.0) {  } 
child {node [b] at (-0.5, 1.0) {  }  
}
child {node [b] at (0.5, 1.0) {  }  
}
;
}}

\newcommand{\forestCE}{
\tikz[planar forest] {

\node [b] at (0.0, 0.0) {  } 
child {node [b] at (0.0, 1.0) {  }  
child {node [b] at (0.0, 1.0) {  }  
}
}
;
}}

\newcommand{\forestDE}{
\tikz[planar forest] {

\node [b] at (0.0, 0.0) {  } 
;
\node [b] at (1.0, 0.0) {  } 
child {node [b] at (0.0, 1.0) {  }  
}
;
}}

\newcommand{\forestEE}{
\tikz[planar forest] {

\node [b] at (0.0, 0.0) {  } 
child {node [b] at (0.0, 1.0) {  }  
}
;
}}

\newcommand{\forestFE}{
\tikz[planar forest] {

\node [b] at (0.0, 0.0) {  } 
;
\node [b] at (1.0, 0.0) {  } 
child {node [b] at (0.0, 1.0) {  }  
}
;
\node [b] at (2.0, 0.0) {  } 
child {node [b] at (0.0, 1.0) {  }  
}
;
}}

\newcommand{\forestGE}{
\tikz[planar forest] {

\node [b] at (0.0, 0.0) {  } 
;
\node [b] at (1.0, 0.0) {  } 
child {node [b] at (-0.5, 1.0) {  }  
child {node [b] at (0.0, 1.0) {  }  
}
}
child {node [b] at (0.5, 1.0) {  }  
}
;
}}

\newcommand{\forestHE}{
\tikz[planar forest] {

\node [b] at (0.0, 0.0) {  } 
;
\node [b] at (1.0, 0.0) {  } 
child {node [b] at (0.0, 1.0) {  }  
child {node [b] at (0.0, 1.0) {  }  
child {node [b] at (0.0, 1.0) {  }  
}
}
}
;
}}

\newcommand{\forestIE}{
\tikz[planar forest] {

\node [b] at (0.0, 0.0) {  } 
child {node [b] at (0.0, 1.0) {  }  
}
;
\node [b] at (1.5, 0.0) {  } 
child {node [b] at (-0.5, 1.0) {  }  
}
child {node [b] at (0.5, 1.0) {  }  
}
;
}}

\newcommand{\forestJE}{
\tikz[planar forest] {

\node [b] at (0.0, 0.0) {  } 
child {node [b] at (0.0, 1.0) {  }  
}
;
\node [b] at (1.0, 0.0) {  } 
child {node [b] at (0.0, 1.0) {  }  
child {node [b] at (0.0, 1.0) {  }  
}
}
;
}}

\newcommand{\forestKE}{
\tikz[planar forest] {

\node [b] at (0.0, 0.0) {  } 
child {node [b] at (-1.0, 1.0) {  }  
}
child {node [b] at (0.0, 1.0) {  }  
child {node [b] at (0.0, 1.0) {  }  
}
}
child {node [b] at (1.0, 1.0) {  }  
}
;
}}

\newcommand{\forestLE}{
\tikz[planar forest] {

\node [b] at (0.0, 0.0) {  } 
child {node [b] at (-0.5, 1.0) {  }  
}
child {node [b] at (0.5, 1.0) {  }  
child {node [b] at (0.0, 1.0) {  }  
child {node [b] at (0.0, 1.0) {  }  
}
}
}
;
}}

\newcommand{\forestME}{
\tikz[planar forest] {

\node [b] at (0.0, 0.0) {  } 
child {node [b] at (-0.5, 1.0) {  }  
child {node [b] at (0.0, 1.0) {  }  
}
}
child {node [b] at (0.5, 1.0) {  }  
child {node [b] at (0.0, 1.0) {  }  
}
}
;
}}

\newcommand{\forestNE}{
\tikz[planar forest] {

\node [b] at (0.0, 0.0) {  } 
child {node [b] at (0.0, 1.0) {  }  
child {node [b] at (-0.5, 1.0) {  }  
}
child {node [b] at (0.5, 1.0) {  }  
child {node [b] at (0.0, 1.0) {  }  
}
}
}
;
}}

\newcommand{\forestOE}{
\tikz[planar forest] {

\node [b] at (0.0, 0.0) {  } 
;
}}

\newcommand{\forestPE}{
\tikz[planar forest] {

\node [b] at (0.0, 0.0) {  } 
;
}}

\newcommand{\forestQE}{
\tikz[planar forest] {

\node [b] at (0.0, 0.0) {  } 
;
}}

\newcommand{\forestRE}{
\tikz[planar forest] {

\node [b] at (0.0, 0.0) {  } 
child {node [b] at (0.0, 1.0) {  }  
}
;
}}

\newcommand{\forestSE}{
\tikz[planar forest] {

\node [b] at (0.0, 0.0) {  } 
child {node [b] at (0.0, 1.0) {  }  
}
;
}}

\newcommand{\forestTE}{
\tikz[planar forest] {

\node [b] at (0.0, 0.0) {  } 
;
}}

\newcommand{\forestUE}{
\tikz[planar forest] {

\node [b] at (0.0, 0.0) {  } 
;
}}

\newcommand{\forestVE}{
\tikz[planar forest] {

\node [b] at (0.0, 0.0) {  } 
child {node [b] at (0.0, 1.0) {  }  
}
;
}}

\newcommand{\forestWE}{
\tikz[planar forest] {

\node [b] at (0.0, 0.0) {  } 
child {node [b] at (0.0, 1.0) {  }  
child {node [b] at (0.0, 1.0) {  }  
}
}
;
}}

\newcommand{\forestXE}{
\tikz[planar forest] {

\node [b] at (0.0, 0.0) {  } 
child {node [b] at (0.0, 1.0) {  }  
child {node [b] at (0.0, 1.0) {  }  
}
}
;
}}

\newcommand{\forestYE}{
\tikz[planar forest] {

\node [b] at (0.0, 0.0) {  } 
child {node [b] at (0.0, 1.0) {  }  
}
;
}}

\newcommand{\forestAF}{
\tikz[planar forest] {

\node [b] at (0.0, 0.0) {  } 
;
}}

\newcommand{\forestBF}{
\tikz[planar forest] {

\node [b] at (0.0, 0.0) {  } 
;
}}

\newcommand{\forestCF}{
\tikz[planar forest] {

\node [b] at (0.0, 0.0) {  } 
child {node [b] at (0.0, 1.0) {  }  
}
;
}}

\newcommand{\forestDF}{
\tikz[planar forest] {

\node [b] at (0.0, 0.0) {  } 
child {node [b] at (0.0, 1.0) {  }  
child {node [b] at (0.0, 1.0) {  }  
}
}
;
}}

\newcommand{\forestEF}{
\tikz[planar forest] {

\node [b] at (0.0, 0.0) {  } 
child {node [b] at (-0.5, 1.0) {  }  
}
child {node [b] at (0.5, 1.0) {  }  
}
;
}}

\newcommand{\forestFF}{
\tikz[planar forest] {

\node [b] at (0.0, 0.0) {  } 
child {node [b] at (-0.5, 1.0) {  }  
}
child {node [b] at (0.5, 1.0) {  }  
}
;
}}

\newcommand{\forestGF}{
\tikz[planar forest] {

\node [b] at (0.0, 0.0) {  } 
;
\node [b] at (1.0, 0.0) {  } 
;
}}

\newcommand{\forestHF}{
\tikz[planar forest] {

\node [b] at (0.0, 0.0) {  } 
;
}}

\newcommand{\forestIF}{
\tikz[planar forest] {

\node [b] at (0.0, 0.0) {  } 
;
}}

\newcommand{\forestJF}{
\tikz[planar forest] {

\node [b] at (0.0, 0.0) {  } 
child {node [b] at (0.0, 1.0) {  }  
}
;
}}

\newcommand{\forestKF}{
\tikz[planar forest] {

\node [b] at (0.0, 0.0) {  } 
child {node [b] at (-0.5, 1.0) {  }  
}
child {node [b] at (0.5, 1.0) {  }  
}
;
}}

\newcommand{\forestLF}{
\tikz[planar forest] {

\node [b] at (0.0, 0.0) {  } 
child {node [b] at (-0.5, 1.0) {  }  
}
child {node [b] at (0.5, 1.0) {  }  
child {node [b] at (0.0, 1.0) {  }  
}
}
;
}}

\newcommand{\forestMF}{
\tikz[planar forest] {

\node [b] at (0.0, 0.0) {  } 
child {node [b] at (-0.5, 1.0) {  }  
}
child {node [b] at (0.5, 1.0) {  }  
child {node [b] at (0.0, 1.0) {  }  
}
}
;
}}

\newcommand{\forestNF}{
\tikz[planar forest] {

\node [b] at (0.0, 0.0) {  } 
;
\node [b] at (1.0, 0.0) {  } 
child {node [b] at (0.0, 1.0) {  }  
}
;
}}

\newcommand{\forestOF}{
\tikz[planar forest] {

\node [b] at (0.0, 0.0) {  } 
;
}}

\newcommand{\forestPF}{
\tikz[planar forest] {

\node [b] at (0.0, 0.0) {  } 
child {node [b] at (0.0, 1.0) {  }  
}
;
}}

\newcommand{\forestQF}{
\tikz[planar forest] {

\node [b] at (0.0, 0.0) {  } 
child {node [b] at (0.0, 1.0) {  }  
}
;
}}

\newcommand{\forestRF}{
\tikz[planar forest] {

\node [b] at (0.0, 0.0) {  } 
;
\node [b] at (1.0, 0.0) {  } 
;
}}

\newcommand{\forestSF}{
\tikz[planar forest] {

\node [b] at (0.0, 0.0) {  } 
child {node [b] at (0.0, 1.0) {  }  
}
;
}}

\newcommand{\forestTF}{
\tikz[planar forest] {

\node [b] at (0.0, 0.0) {  } 
;
}}

\newcommand{\forestUF}{
\tikz[planar forest] {

\node [b] at (0.0, 0.0) {  } 
child {node [b] at (-0.5, 1.0) {  }  
}
child {node [b] at (0.5, 1.0) {  }  
}
;
}}

\newcommand{\forestVF}{
\tikz[planar forest] {

\node [b] at (0.0, 0.0) {  } 
;
}}

\newcommand{\forestWF}{
\tikz[planar forest] {

\node [b] at (0.0, 0.0) {  } 
child {node [b] at (0.0, 1.0) {  }  
child {node [b] at (0.0, 1.0) {  }  
}
}
;
}}

\newcommand{\forestXF}{
\tikz[planar forest] {

\node [b] at (0.0, 0.0) {  } 
child {node [b] at (-0.5, 1.0) {  }  
}
child {node [b] at (0.5, 1.0) {  }  
child {node [b] at (0.0, 1.0) {  }  
}
}
;
}}

\newcommand{\forestYF}{
\tikz[planar forest] {

\node [b] at (0.0, 0.0) {  } 
child {node [b] at (-0.5, 1.0) {  }  
}
child {node [b] at (0.5, 1.0) {  }  
}
;
}}

\newcommand{\forestAG}{
\tikz[planar forest] {

\node [b] at (0.0, 0.0) {  } 
;
}}

\newcommand{\forestBG}{
\tikz[planar forest] {

\node [b] at (0.0, 0.0) {  } 
child {node [b] at (0.0, 1.0) {  }  
}
;
}}

\newcommand{\forestCG}{
\tikz[planar forest] {

\node [b] at (0.0, 0.0) {  } 
;
}}

\newcommand{\forestDG}{
\tikz[planar forest] {

\node [b] at (0.0, 0.0) {  } 
child {node [b] at (0.0, 1.0) {  }  
}
;
}}

\newcommand{\forestEG}{
\tikz[planar forest] {

\node [b] at (0.0, 0.0) {  } 
child {node [b] at (-0.5, 1.0) {  }  
}
child {node [b] at (0.5, 1.0) {  }  
}
;
}}

\newcommand{\forestFG}{
\tikz[planar forest] {

\node [b] at (0.0, 0.0) {  } 
;
}}

\newcommand{\forestGG}{
\tikz[planar forest] {

\node [b] at (0.0, 0.0) {  } 
;
}}

\newcommand{\forestHG}{
\tikz[planar forest] {

\node [b] at (0.0, 0.0) {  } 
;
}}

\newcommand{\forestIG}{
\tikz[planar forest] {

\node [b] at (0.0, 0.0) {  } 
child {node [b] at (0.0, 1.0) {  }  
}
;
}}

\newcommand{\forestJG}{
\tikz[planar forest] {

\node [b] at (0.0, 0.0) {  } 
child {node [b] at (0.0, 1.0) {  }  
}
;
}}

\newcommand{\forestKG}{
\tikz[planar forest] {

\node [b] at (0.0, 0.0) {  } 
;
}}

\newcommand{\forestLG}{
\tikz[planar forest] {

\node [b] at (0.0, 0.0) {  } 
;
}}

\newcommand{\forestMG}{
\tikz[planar forest] {

\node [b] at (0.0, 0.0) {  } 
child {node [b] at (0.0, 1.0) {  }  
}
;
}}

\newcommand{\forestNG}{
\tikz[planar forest] {

\node [b] at (0.0, 0.0) {  } 
child {node [b] at (0.0, 1.0) {  }  
child {node [b] at (0.0, 1.0) {  }  
}
}
;
}}

\newcommand{\forestOG}{
\tikz[planar forest] {

\node [b] at (0.0, 0.0) {  } 
child {node [b] at (0.0, 1.0) {  }  
child {node [b] at (0.0, 1.0) {  }  
}
}
;
}}

\newcommand{\forestPG}{
\tikz[planar forest] {

\node [b] at (0.0, 0.0) {  } 
child {node [b] at (0.0, 1.0) {  }  
}
;
}}

\newcommand{\forestQG}{
\tikz[planar forest] {

\node [b] at (0.0, 0.0) {  } 
;
}}

\newcommand{\forestRG}{
\tikz[planar forest] {

\node [b] at (0.0, 0.0) {  } 
;
}}

\newcommand{\forestSG}{
\tikz[planar forest] {

\node [b] at (0.0, 0.0) {  } 
child {node [b] at (0.0, 1.0) {  }  
}
;
}}

\newcommand{\forestTG}{
\tikz[planar forest] {

\node [b] at (0.0, 0.0) {  } 
child {node [b] at (0.0, 1.0) {  }  
child {node [b] at (0.0, 1.0) {  }  
}
}
;
}}

\newcommand{\forestUG}{
\tikz[planar forest] {

\node [b] at (0.0, 0.0) {  } 
child {node [b] at (-0.5, 1.0) {  }  
}
child {node [b] at (0.5, 1.0) {  }  
}
;
}}

\newcommand{\forestVG}{
\tikz[planar forest] {

\node [b] at (0.0, 0.0) {  } 
child {node [b] at (-0.5, 1.0) {  }  
}
child {node [b] at (0.5, 1.0) {  }  
}
;
}}

\newcommand{\forestWG}{
\tikz[planar forest] {

\node [b] at (0.0, 0.0) {  } 
;
}}

\newcommand{\forestXG}{
\tikz[planar forest] {

\node [b] at (0.0, 0.0) {  } 
;
}}

\newcommand{\forestYG}{
\tikz[planar forest] {

\node [b] at (0.0, 0.0) {  } 
;
}}

\newcommand{\forestAH}{
\tikz[planar forest] {

\node [b] at (0.0, 0.0) {  } 
child {node [b] at (0.0, 1.0) {  }  
}
;
}}

\newcommand{\forestBH}{
\tikz[planar forest] {

\node [b] at (0.0, 0.0) {  } 
child {node [b] at (-0.5, 1.0) {  }  
}
child {node [b] at (0.5, 1.0) {  }  
}
;
}}

\newcommand{\forestCH}{
\tikz[planar forest] {

\node [b] at (0.0, 0.0) {  } 
child {node [b] at (0.0, 1.0) {  }  
child {node [b] at (0.0, 1.0) {  }  
}
}
;
}}

\newcommand{\forestDH}{
\tikz[planar forest] {

\node [b] at (0.0, 0.0) {  } 
child {node [b] at (-0.5, 1.0) {  }  
}
child {node [b] at (0.5, 1.0) {  }  
}
;
}}

\newcommand{\forestEH}{
\tikz[planar forest] {

\node [b] at (0.0, 0.0) {  } 
child {node [b] at (-1.0, 1.0) {  }  
}
child {node [b] at (0.0, 1.0) {  }  
}
child {node [b] at (1.0, 1.0) {  }  
}
;
}}

\newcommand{\forestFH}{
\tikz[planar forest] {

\node [b] at (0.0, 0.0) {  } 
child {node [b] at (0.0, 1.0) {  }  
child {node [b] at (-0.5, 1.0) {  }  
}
child {node [b] at (0.5, 1.0) {  }  
}
}
;
}}

\newcommand{\forestGH}{
\tikz[planar forest] {

\node [b] at (0.0, 0.0) {  } 
child {node [b] at (-1.0, 1.0) {  }  
}
child {node [b] at (0.0, 1.0) {  }  
}
child {node [b] at (1.0, 1.0) {  }  
child {node [b] at (0.0, 1.0) {  }  
}
}
;
}}

\newcommand{\forestHH}{
\tikz[planar forest] {

\node [b] at (0.0, 0.0) {  } 
child {node [b] at (0.0, 1.0) {  }  
child {node [b] at (-0.5, 1.0) {  }  
}
child {node [b] at (0.5, 1.0) {  }  
child {node [b] at (0.0, 1.0) {  }  
}
}
}
;
}}

\newcommand{\forestIH}{
\tikz[planar forest] {

\node [b] at (0.0, 0.0) {  } 
child {node [b] at (-0.5, 1.0) {  }  
child {node [b] at (-0.5, 1.0) {  }  
}
child {node [b] at (0.5, 1.0) {  }  
}
}
child {node [b] at (0.5, 1.0) {  }  
}
;
}}

\newcommand{\forestJH}{
\tikz[planar forest] {

\node [b] at (0.0, 0.0) {  } 
child {node [b] at (0.0, 1.0) {  }  
child {node [b] at (0.0, 1.0) {  }  
child {node [b] at (-0.5, 1.0) {  }  
}
child {node [b] at (0.5, 1.0) {  }  
}
}
}
;
}}

\newcommand{\forestKH}{
\tikz[planar forest] {

\node [b] at (0.0, 0.0) {  } 
child {node [b] at (0.0, 1.0) {  }  
}
;
\node [b] at (1.0, 0.0) {  } 
;
}}

\newcommand{\forestLH}{
\tikz[planar forest] {

\node [b] at (0.0, 0.0) {  } 
;
\node [b] at (1.0, 0.0) {  } 
child {node [b] at (0.0, 1.0) {  }  
}
;
}}

\newcommand{\forestMH}{
\tikz[planar forest] {

\node [b] at (0.0, 0.0) {  } 
child {node [b] at (-0.5, 1.0) {  }  
}
child {node [b] at (0.5, 1.0) {  }  
child {node [b] at (0.0, 1.0) {  }  
}
}
;
}}

\newcommand{\forestNH}{
\tikz[planar forest] {

\node [b] at (0.0, 0.0) {  } 
child {node [b] at (-0.5, 1.0) {  }  
child {node [b] at (0.0, 1.0) {  }  
}
}
child {node [b] at (0.5, 1.0) {  }  
}
;
}}

\newcommand{\forestOH}{
\tikz[planar forest] {

\node [b] at (0.0, 0.0) {  } 
child {node [l] at (0.0, 1.0) { $\times$ }  
}
;
}}

\newcommand{\forestPH}{
\tikz[planar forest] {

\node [b] at (0.0, 0.0) {  } 
child {node [b] at (-0.5, 1.0) {  }  
child {node [l] at (-0.5, 1.0) { $\times$ }  
}
child {node [l] at (0.5, 1.0) { $\times$ }  
}
}
child {node [l] at (0.5, 1.0) { $\times$ }  
}
;
}}

\newcommand{\forestQH}{
\tikz[planar forest] {

\node [b] at (0.0, 0.0) {  } 
child {node [b] at (-0.5, 1.0) {  }  
child {node [l] at (0.0, 1.0) { $\times$ }  
}
}
child {node [b] at (0.5, 1.0) {  }  
child {node [l] at (0.0, 1.0) { $\times$ }  
}
}
;
}}

\newcommand{\forestRH}{
\tikz[planar forest] {

\node [b] at (0.0, 0.0) {  } 
child {node [b] at (-0.5, 1.0) {  }  
child {node [lc] at (0.0, 1.0) { 1 }  
}
}
child {node [b] at (0.5, 1.0) {  }  
child {node [lc] at (0.0, 1.0) { 1 }  
}
}
;
}}

\newcommand{\forestSH}{
\tikz[planar forest] {

\node [b] at (0.0, 0.0) {  } 
child {node [b] at (-1.5, 1.0) {  }  
child {node [lc] at (-0.5, 1.0) { 1 }  
}
child {node [lc] at (0.5, 1.0) { 2 }  
}
}
child {node [b] at (0.0, 1.0) {  }  
child {node [lc] at (0.0, 1.0) { 1 }  
}
}
child {node [b] at (1.5, 1.0) {  }  
child {node [b] at (-0.5, 1.0) {  }  
}
child {node [lc] at (0.5, 1.0) { 2 }  
}
}
;
}}

\newcommand{\forestTH}{
\tikz[planar forest] {

\node [b] at (0.0, 0.0) {  } 
child {node [lc] at (-1.5, 1.0) { 1 }  
}
child {node [lc] at (-0.5, 1.0) { 1 }  
}
child {node [lc] at (0.5, 1.0) { 2 }  
}
child {node [b] at (1.5, 1.0) {  }  
child {node [lc] at (0.0, 1.0) { 2 }  
}
}
;
}}

\newcommand{\forestUH}{
\tikz[planar forest] {

\node [b] at (0.0, 0.0) {  } 
child {node [lc] at (-1.5, 1.0) { 1 }  
}
child {node [lc] at (-0.5, 1.0) { 1 }  
}
child {node [lc] at (0.5, 1.0) { 2 }  
}
child {node [b] at (1.5, 1.0) {  }  
child {node [lc] at (0.0, 1.0) { 2 }  
}
}
;
}}

\newcommand{\forestVH}{
\tikz[planar forest] {

\node [b] at (0.0, 0.0) {  } 
child {node [lc] at (-1.5, 1.0) { 2 }  
}
child {node [lc] at (-0.5, 1.0) { 2 }  
}
child {node [lc] at (0.5, 1.0) { 1 }  
}
child {node [b] at (1.5, 1.0) {  }  
child {node [lc] at (0.0, 1.0) { 1 }  
}
}
;
}}

\newcommand{\forestWH}{
\tikz[planar forest] {

\draw (0.0,1.0) arc (90:270:0.5);
\draw (0.0,1.0) arc (90:-90:0.5);
\draw (0.0,0.0) edge (0.0,0.0);
\node [draw=none] at (0.0, 0.0) { } 
child {node [b] at (0.0, 1.0) {  } edge from parent[draw=none] 
}
;
\node [b] at (1.1, 0.0) {  } 
;
}}

\newcommand{\forestXH}{
\tikz[planar forest] {

\draw (0.0,1.0) arc (90:270:0.5);
\draw (0.0,1.0) arc (90:-90:0.5);
\draw (0.0,0.0) edge (0.0,0.0);
\node [draw=none] at (0.0, 0.0) { } 
child {node [b] at (0.0, 1.0) {  } edge from parent[draw=none] 
}
;
\draw (1.2000000000000002,1.0) arc (90:270:0.5);
\draw (1.2000000000000002,1.0) arc (90:-90:0.5);
\draw (1.2000000000000002,0.0) edge (1.2000000000000002,0.0);
\node [draw=none] at (1.2000000000000002, 0.0) { } 
child {node [b] at (0.0, 1.0) {  } edge from parent[draw=none] 
child {node [b] at (0.0, 1.0) {  }  
}
}
;
\node [b] at (2.7, 0.0) {  } 
child {node [b] at (-0.5, 1.0) {  }  
}
child {node [b] at (0.5, 1.0) {  }  
}
;
}}

\newcommand{\forestYH}{
\tikz[planar forest] {

\draw (-1.0,1.0) edge (0.0,1.0);
\draw (0.0,1.0) edge (1.5000000000000004,1.0);
\draw (-1.0,1.0) arc (90:270:0.5);
\draw (1.5000000000000004,1.0) arc (90:-90:0.5);
\draw (-1.0,0.0) edge (1.5000000000000004,0.0);
\node [draw=none] at (0.0, 0.0) { } 
child {node [b] at (-1.0, 1.0) {  } edge from parent[draw=none] 
}
child {node [b] at (0.0, 1.0) {  } edge from parent[draw=none] 
child {node [b] at (-0.5, 1.0) {  }  
}
child {node [b] at (0.5000000000000004, 1.0) {  }  
}
}
child {node [b] at (1.5000000000000004, 1.0) {  } edge from parent[draw=none] 
child {node [b] at (0.0, 1.0) {  }  
}
}
;
\draw (2.7,1.0) arc (90:270:0.5);
\draw (2.7,1.0) arc (90:-90:0.5);
\draw (2.7,0.0) edge (2.7,0.0);
\node [draw=none] at (2.7, 0.0) { } 
child {node [b] at (0.0, 1.0) {  } edge from parent[draw=none] 
}
;
\node [b] at (3.8000000000000003, 0.0) {  } 
child {node [b] at (0.0, 1.0) {  }  
child {node [b] at (-0.5, 1.0) {  }  
}
child {node [b] at (0.5000000000000004, 1.0) {  }  
}
}
;
}}

\newcommand{\forestAI}{
\tikz[planar forest] {

\node [lc] at (0.0, 0.0) { 1 } 
;
}}

\newcommand{\forestBI}{
\tikz[planar forest] {

\node [lc] at (0.0, 0.0) { 1 } 
child {node [lc,label={ [label distance=-1mm]180:{ \scriptsize 1 } }] at (-0.5, 1.0) { 2 }  
}
child {node [lc,label={ [label distance=-1mm]0:{ \scriptsize 2 } }] at (0.5, 1.0) { 3 }  
}
;
}}

\newcommand{\forestCI}{
\tikz[planar forest] {

\node [lc] at (0.0, 0.0) { 1 } 
child {node [lc,label={ [label distance=-1mm]180:{ \scriptsize 1 } }] at (-0.5, 1.0) { 1 }  
}
child {node [lc,label={ [label distance=-1mm]0:{ \scriptsize 4 } }] at (0.5, 1.0) { 2 }  
child {node [lc,label={ [label distance=-1mm]180:{ \scriptsize 2 } }] at (-0.5, 1.0) { 3 }  
}
child {node [lc,label={ [label distance=-1mm]0:{ \scriptsize 3 } }] at (0.5, 1.0) { 2 }  
}
}
;
}}

\newcommand{\forestDI}{
\tikz[planar forest] {

\draw (0.0,1.0) arc (90:270:0.5);
\draw (0.0,1.0) arc (90:-90:0.5);
\draw (0.0,0.0) edge (0.0,0.0);
\node [draw=none] at (0.0, 0.0) { } 
child {node [b] at (0.0, 1.0) {  } edge from parent[draw=none] 
}
;
}}

\newcommand{\forestEI}{
\tikz[planar forest] {

\draw (0.0,-0.1) edge (1.0,-0.1);
\draw (0.0,0.1) edge (1.0,0.1);
;
\node [b] at (0.0, 0.0) {  } 
;
\node [b] at (1.0, 0.0) {  } 
;
}}

\newcommand{\forestFI}{
\tikz[planar forest] {

\node [b] at (0.0, 0.0) {  } 
;
}}

\newcommand{\forestGI}{
\tikz[planar forest] {

\draw (0.0,1.0) arc (90:270:0.5);
\draw (0.0,1.0) arc (90:-90:0.5);
\draw (0.0,0.0) edge (0.0,0.0);
\node [draw=none] at (0.0, 0.0) { } 
child {node [b] at (0.0, 1.0) {  } edge from parent[draw=none] 
child {node [b] at (0.0, 1.0) {  }  
}
}
;
}}

\newcommand{\forestHI}{
\tikz[planar forest] {

\draw (-0.5,1.0) edge (0.5,1.0);
\draw (-0.5,1.0) arc (90:270:0.5);
\draw (0.5,1.0) arc (90:-90:0.5);
\draw (-0.5,0.0) edge (0.5,0.0);
\node [draw=none] at (0.0, 0.0) { } 
child {node [b] at (-0.5, 1.0) {  } edge from parent[draw=none] 
}
child {node [b] at (0.5, 1.0) {  } edge from parent[draw=none] 
}
;
}}

\newcommand{\forestII}{
\tikz[planar forest] {

\draw (0.0,1.0) arc (90:270:0.5);
\draw (0.0,1.0) arc (90:-90:0.5);
\draw (0.0,0.0) edge (0.0,0.0);
\node [draw=none] at (0.0, 0.0) { } 
child {node [b] at (0.0, 1.0) {  } edge from parent[draw=none] 
child {node [lc] at (-0.5, 1.0) { 1 }  
}
child {node [lc] at (0.5, 1.0) { 1 }  
}
}
;
}}

\newcommand{\forestJI}{
\tikz[planar forest] {

\draw (0.0,-0.1) edge (1.0,-0.1);
\draw (0.0,0.1) edge (1.0,0.1);
;
\node [b] at (0.0, 0.0) {  } 
;
\node [b] at (1.0, 0.0) {  } 
child {node [b] at (0.0, 1.0) {  }  
}
;
}}

\newcommand{\forestKI}{
\tikz[planar forest] {

\draw (0.0,-0.1) edge (1.0,-0.1);
\draw (0.0,0.1) edge (1.0,0.1);
;
\node [b] at (0.0, 0.0) {  } 
;
\node [b] at (1.0, 0.0) {  } 
child {node [lc] at (-0.5, 1.0) { 1 }  
}
child {node [lc] at (0.5, 1.0) { 1 }  
}
;
}}

\newcommand{\forestLI}{
\tikz[planar forest] {

\draw (0.0,-0.1) edge (1.0,-0.1);
\draw (0.0,0.1) edge (1.0,0.1);
;
\node [b] at (0.0, 0.0) {  } 
child {node [lc] at (0.0, 1.0) { 1 }  
}
;
\node [b] at (1.0, 0.0) {  } 
child {node [lc] at (0.0, 1.0) { 1 }  
}
;
}}

\newcommand{\forestMI}{
\tikz[planar forest] {

\node [b] at (0.0, 0.0) {  } 
child {node [b] at (0.0, 1.0) {  }  
}
;
}}

\newcommand{\forestNI}{
\tikz[planar forest] {

\node [b] at (0.0, 0.0) {  } 
child {node [lc] at (-0.5, 1.0) { 1 }  
}
child {node [lc] at (0.5, 1.0) { 1 }  
}
;
}}

\newcommand{\forestOI}{
\tikz[planar forest] {

\draw (0.0,1.0) arc (90:270:0.5);
\draw (0.0,1.0) arc (90:-90:0.5);
\draw (0.0,0.0) edge (0.0,0.0);
\node [draw=none] at (0.0, 0.0) { } 
child {node [b] at (0.0, 1.0) {  } edge from parent[draw=none] 
child {node [lc] at (0.0, 1.0) { 1 }  
}
}
;
\node [lc] at (1.1, 0.0) { 1 } 
;
}}

\newcommand{\forestPI}{
\tikz[planar forest] {

\draw (0.0,-0.1) edge (1.0,-0.1);
\draw (0.0,0.1) edge (1.0,0.1);
;
\node [b] at (0.0, 0.0) {  } 
;
\node [b] at (1.0, 0.0) {  } 
child {node [lc] at (0.0, 1.0) { 1 }  
}
;
\node [lc] at (2.0, 0.0) { 1 } 
;
}}

\newcommand{\forestQI}{
\tikz[planar forest] {

\node [b] at (0.0, 0.0) {  } 
child {node [lc] at (0.0, 1.0) { 1 }  
}
;
\node [lc] at (1.0, 0.0) { 1 } 
;
}}

\newcommand{\forestRI}{
\tikz[planar forest] {

\draw (0.0,1.0) arc (90:270:0.5);
\draw (0.0,1.0) arc (90:-90:0.5);
\draw (0.0,0.0) edge (0.0,0.0);
\node [draw=none] at (0.0, 0.0) { } 
child {node [b] at (0.0, 1.0) {  } edge from parent[draw=none] 
child {node [b] at (0.0, 1.0) {  }  
child {node [b] at (0.0, 1.0) {  }  
}
}
}
;
}}

\newcommand{\forestSI}{
\tikz[planar forest] {

\draw (0.0,1.0) arc (90:270:0.5);
\draw (0.0,1.0) arc (90:-90:0.5);
\draw (0.0,0.0) edge (0.0,0.0);
\node [draw=none] at (0.0, 0.0) { } 
child {node [b] at (0.0, 1.0) {  } edge from parent[draw=none] 
child {node [b] at (-0.5, 1.0) {  }  
}
child {node [b] at (0.5, 1.0) {  }  
}
}
;
}}

\newcommand{\forestTI}{
\tikz[planar forest] {

\draw (-0.5,1.0) edge (0.5,1.0);
\draw (-0.5,1.0) arc (90:270:0.5);
\draw (0.5,1.0) arc (90:-90:0.5);
\draw (-0.5,0.0) edge (0.5,0.0);
\node [draw=none] at (0.0, 0.0) { } 
child {node [b] at (-0.5, 1.0) {  } edge from parent[draw=none] 
child {node [b] at (0.0, 1.0) {  }  
}
}
child {node [b] at (0.5, 1.0) {  } edge from parent[draw=none] 
}
;
}}

\newcommand{\forestUI}{
\tikz[planar forest] {

\draw (-1.0,1.0) edge (0.0,1.0);
\draw (0.0,1.0) edge (1.0,1.0);
\draw (-1.0,1.0) arc (90:270:0.5);
\draw (1.0,1.0) arc (90:-90:0.5);
\draw (-1.0,0.0) edge (1.0,0.0);
\node [draw=none] at (0.0, 0.0) { } 
child {node [b] at (-1.0, 1.0) {  } edge from parent[draw=none] 
}
child {node [b] at (0.0, 1.0) {  } edge from parent[draw=none] 
}
child {node [b] at (1.0, 1.0) {  } edge from parent[draw=none] 
}
;
}}

\newcommand{\forestVI}{
\tikz[planar forest] {

\draw (0.0,1.0) arc (90:270:0.5);
\draw (0.0,1.0) arc (90:-90:0.5);
\draw (0.0,0.0) edge (0.0,0.0);
\node [draw=none] at (0.0, 0.0) { } 
child {node [b] at (0.0, 1.0) {  } edge from parent[draw=none] 
child {node [b] at (0.0, 1.0) {  }  
child {node [lc] at (-0.5, 1.0) { 1 }  
}
child {node [lc] at (0.5, 1.0) { 1 }  
}
}
}
;
}}

\newcommand{\forestWI}{
\tikz[planar forest] {

\draw (0.0,1.0) arc (90:270:0.5);
\draw (0.0,1.0) arc (90:-90:0.5);
\draw (0.0,0.0) edge (0.0,0.0);
\node [draw=none] at (0.0, 0.0) { } 
child {node [b] at (0.0, 1.0) {  } edge from parent[draw=none] 
child {node [b] at (-0.5, 1.0) {  }  
child {node [lc] at (0.0, 1.0) { 1 }  
}
}
child {node [lc] at (0.5, 1.0) { 1 }  
}
}
;
}}

\newcommand{\forestXI}{
\tikz[planar forest] {

\draw (0.0,1.0) arc (90:270:0.5);
\draw (0.0,1.0) arc (90:-90:0.5);
\draw (0.0,0.0) edge (0.0,0.0);
\node [draw=none] at (0.0, 0.0) { } 
child {node [b] at (0.0, 1.0) {  } edge from parent[draw=none] 
child {node [b] at (-1.0, 1.0) {  }  
}
child {node [lc] at (0.0, 1.0) { 1 }  
}
child {node [lc] at (1.0, 1.0) { 1 }  
}
}
;
}}

\newcommand{\forestYI}{
\tikz[planar forest] {

\draw (-2.220446049250313e-16,1.0) arc (90:270:0.5);
\draw (-2.220446049250313e-16,1.0) arc (90:-90:0.5);
\draw (-2.220446049250313e-16,0.0) edge (-2.220446049250313e-16,0.0);
\node [draw=none] at (-2.220446049250313e-16, 0.0) { } 
child {node [b] at (0.0, 1.0) {  } edge from parent[draw=none] 
child {node [lc] at (0.0, 1.0) { 1 }  
}
}
;
\draw (1.2,1.0) arc (90:270:0.5);
\draw (1.2,1.0) arc (90:-90:0.5);
\draw (1.2,0.0) edge (1.2,0.0);
\node [draw=none] at (1.2, 0.0) { } 
child {node [b] at (0.0, 1.0) {  } edge from parent[draw=none] 
child {node [lc] at (0.0, 1.0) { 1 }  
}
}
;
}}

\newcommand{\forestAJ}{
\tikz[planar forest] {

\draw (-0.5,1.0) edge (0.5,1.0);
\draw (-0.5,1.0) arc (90:270:0.5);
\draw (0.5,1.0) arc (90:-90:0.5);
\draw (-0.5,0.0) edge (0.5,0.0);
\node [draw=none] at (0.0, 0.0) { } 
child {node [b] at (-0.5, 1.0) {  } edge from parent[draw=none] 
child {node [lc] at (0.0, 1.0) { 1 }  
}
}
child {node [b] at (0.5, 1.0) {  } edge from parent[draw=none] 
child {node [lc] at (0.0, 1.0) { 1 }  
}
}
;
}}

\newcommand{\forestBJ}{
\tikz[planar forest] {

\draw (-0.5,1.0) edge (0.5,1.0);
\draw (-0.5,1.0) arc (90:270:0.5);
\draw (0.5,1.0) arc (90:-90:0.5);
\draw (-0.5,0.0) edge (0.5,0.0);
\node [draw=none] at (0.0, 0.0) { } 
child {node [b] at (-0.5, 1.0) {  } edge from parent[draw=none] 
child {node [lc] at (-0.5, 1.0) { 1 }  
}
child {node [lc] at (0.5, 1.0) { 1 }  
}
}
child {node [b] at (0.5, 1.0) {  } edge from parent[draw=none] 
}
;
}}

\newcommand{\forestCJ}{
\tikz[planar forest] {

\draw (0.0,1.0) arc (90:270:0.5);
\draw (0.0,1.0) arc (90:-90:0.5);
\draw (0.0,0.0) edge (0.0,0.0);
\node [draw=none] at (0.0, 0.0) { } 
child {node [b] at (0.0, 1.0) {  } edge from parent[draw=none] 
child {node [lc] at (-1.5, 1.0) { 1 }  
}
child {node [lc] at (-0.5, 1.0) { 1 }  
}
child {node [lc] at (0.5, 1.0) { 2 }  
}
child {node [lc] at (1.5, 1.0) { 2 }  
}
}
;
}}

\newcommand{\forestDJ}{
\tikz[planar forest] {

\draw (0.0,-0.1) edge (1.0,-0.1);
\draw (0.0,0.1) edge (1.0,0.1);
;
\node [b] at (0.0, 0.0) {  } 
;
\node [b] at (1.0, 0.0) {  } 
child {node [b] at (0.0, 1.0) {  }  
child {node [b] at (0.0, 1.0) {  }  
}
}
;
}}

\newcommand{\forestEJ}{
\tikz[planar forest] {

\draw (0.0,-0.1) edge (1.0,-0.1);
\draw (0.0,0.1) edge (1.0,0.1);
;
\node [b] at (0.0, 0.0) {  } 
;
\node [b] at (1.0, 0.0) {  } 
child {node [b] at (-0.5, 1.0) {  }  
}
child {node [b] at (0.5, 1.0) {  }  
}
;
}}

\newcommand{\forestFJ}{
\tikz[planar forest] {

\draw (0.0,-0.1) edge (1.0,-0.1);
\draw (0.0,0.1) edge (1.0,0.1);
;
\node [b] at (0.0, 0.0) {  } 
child {node [b] at (0.0, 1.0) {  }  
}
;
\node [b] at (1.0, 0.0) {  } 
child {node [b] at (0.0, 1.0) {  }  
}
;
}}

\newcommand{\forestGJ}{
\tikz[planar forest] {

\draw (0.0,-0.1) edge (1.0,-0.1);
\draw (0.0,0.1) edge (1.0,0.1);
;
\node [b] at (0.0, 0.0) {  } 
;
\node [b] at (1.0, 0.0) {  } 
child {node [b] at (0.0, 1.0) {  }  
child {node [lc] at (-0.5, 1.0) { 1 }  
}
child {node [lc] at (0.5, 1.0) { 1 }  
}
}
;
}}

\newcommand{\forestHJ}{
\tikz[planar forest] {

\draw (0.0,-0.1) edge (1.0,-0.1);
\draw (0.0,0.1) edge (1.0,0.1);
;
\node [b] at (0.0, 0.0) {  } 
;
\node [b] at (1.0, 0.0) {  } 
child {node [b] at (-0.5, 1.0) {  }  
child {node [lc] at (0.0, 1.0) { 1 }  
}
}
child {node [lc] at (0.5, 1.0) { 1 }  
}
;
}}

\newcommand{\forestIJ}{
\tikz[planar forest] {

\draw (0.0,-0.1) edge (1.0,-0.1);
\draw (0.0,0.1) edge (1.0,0.1);
;
\node [b] at (0.0, 0.0) {  } 
child {node [lc] at (0.0, 1.0) { 1 }  
}
;
\node [b] at (1.0, 0.0) {  } 
child {node [b] at (0.0, 1.0) {  }  
child {node [lc] at (0.0, 1.0) { 1 }  
}
}
;
}}

\newcommand{\forestJJ}{
\tikz[planar forest] {

\draw (0.0,-0.1) edge (1.0,-0.1);
\draw (0.0,0.1) edge (1.0,0.1);
;
\node [b] at (0.0, 0.0) {  } 
;
\node [b] at (1.0, 0.0) {  } 
child {node [b] at (-1.0, 1.0) {  }  
}
child {node [lc] at (0.0, 1.0) { 1 }  
}
child {node [lc] at (1.0, 1.0) { 1 }  
}
;
}}

\newcommand{\forestKJ}{
\tikz[planar forest] {

\draw (0.0,-0.1) edge (1.5,-0.1);
\draw (0.0,0.1) edge (1.5,0.1);
;
\node [b] at (0.0, 0.0) {  } 
child {node [lc] at (0.0, 1.0) { 1 }  
}
;
\node [b] at (1.5, 0.0) {  } 
child {node [b] at (-0.5, 1.0) {  }  
}
child {node [lc] at (0.5, 1.0) { 1 }  
}
;
}}

\newcommand{\forestLJ}{
\tikz[planar forest] {

\draw (0.0,-0.1) edge (1.5,-0.1);
\draw (0.0,0.1) edge (1.5,0.1);
;
\node [b] at (0.0, 0.0) {  } 
child {node [lc] at (-0.5, 1.0) { 1 }  
}
child {node [lc] at (0.5, 1.0) { 1 }  
}
;
\node [b] at (1.5, 0.0) {  } 
child {node [b] at (0.0, 1.0) {  }  
}
;
}}

\newcommand{\forestMJ}{
\tikz[planar forest] {

\draw (0.0,-0.1) edge (1.0,-0.1);
\draw (0.0,0.1) edge (1.0,0.1);
;
\node [b] at (0.0, 0.0) {  } 
;
\node [b] at (1.0, 0.0) {  } 
child {node [lc] at (-1.5, 1.0) { 1 }  
}
child {node [lc] at (-0.5, 1.0) { 1 }  
}
child {node [lc] at (0.5, 1.0) { 2 }  
}
child {node [lc] at (1.5, 1.0) { 2 }  
}
;
}}

\newcommand{\forestNJ}{
\tikz[planar forest] {

\draw (0.0,-0.1) edge (2.0,-0.1);
\draw (0.0,0.1) edge (2.0,0.1);
;
\node [b] at (0.0, 0.0) {  } 
child {node [lc] at (0.0, 1.0) { 2 }  
}
;
\node [b] at (2.0, 0.0) {  } 
child {node [lc] at (-1.0, 1.0) { 1 }  
}
child {node [lc] at (0.0, 1.0) { 1 }  
}
child {node [lc] at (1.0, 1.0) { 2 }  
}
;
}}

\newcommand{\forestOJ}{
\tikz[planar forest] {

\draw (0.0,-0.1) edge (2.0,-0.1);
\draw (0.0,0.1) edge (2.0,0.1);
;
\node [b] at (0.0, 0.0) {  } 
child {node [lc] at (-0.5, 1.0) { 2 }  
}
child {node [lc] at (0.5, 1.0) { 2 }  
}
;
\node [b] at (2.0, 0.0) {  } 
child {node [lc] at (-0.5, 1.0) { 1 }  
}
child {node [lc] at (0.5, 1.0) { 1 }  
}
;
}}

\newcommand{\forestPJ}{
\tikz[planar forest] {

\draw (0.0,-0.1) edge (2.0,-0.1);
\draw (0.0,0.1) edge (2.0,0.1);
;
\node [b] at (0.0, 0.0) {  } 
child {node [lc] at (-0.5, 1.0) { 1 }  
}
child {node [lc] at (0.5, 1.0) { 2 }  
}
;
\node [b] at (2.0, 0.0) {  } 
child {node [lc] at (-0.5, 1.0) { 1 }  
}
child {node [lc] at (0.5, 1.0) { 2 }  
}
;
}}

\newcommand{\forestQJ}{
\tikz[planar forest] {

\draw (0.0,-0.1) edge (1.0,-0.1);
\draw (0.0,0.1) edge (1.0,0.1);
;
\node [b] at (0.0, 0.0) {  } 
;
\node [b] at (1.0, 0.0) {  } 
child {node [lc] at (0.0, 1.0) { 1 }  
}
;
\draw (2.1,1.0) arc (90:270:0.5);
\draw (2.1,1.0) arc (90:-90:0.5);
\draw (2.1,0.0) edge (2.1,0.0);
\node [draw=none] at (2.1, 0.0) { } 
child {node [b] at (0.0, 1.0) {  } edge from parent[draw=none] 
child {node [lc] at (0.0, 1.0) { 1 }  
}
}
;
}}

\newcommand{\forestRJ}{
\tikz[planar forest] {

\draw (0.0,-0.1) edge (1.0,-0.1);
\draw (0.0,0.1) edge (1.0,0.1);
;
\draw (2.0,-0.1) edge (3.0,-0.1);
\draw (2.0,0.1) edge (3.0,0.1);
;
\node [b] at (0.0, 0.0) {  } 
;
\node [b] at (1.0, 0.0) {  } 
child {node [lc] at (0.0, 1.0) { 1 }  
}
;
\node [b] at (2.0, 0.0) {  } 
child {node [lc] at (0.0, 1.0) { 1 }  
}
;
\node [b] at (3.0, 0.0) {  } 
;
}}

\newcommand{\forestSJ}{
\tikz[planar forest] {

\node [b] at (0.0, 0.0) {  } 
child {node [b] at (0.0, 1.0) {  }  
child {node [b] at (0.0, 1.0) {  }  
}
}
;
}}

\newcommand{\forestTJ}{
\tikz[planar forest] {

\node [b] at (0.0, 0.0) {  } 
child {node [b] at (-0.5, 1.0) {  }  
}
child {node [b] at (0.5, 1.0) {  }  
}
;
}}

\newcommand{\forestUJ}{
\tikz[planar forest] {

\node [b] at (0.0, 0.0) {  } 
child {node [b] at (0.0, 1.0) {  }  
child {node [lc] at (-0.5, 1.0) { 1 }  
}
child {node [lc] at (0.5, 1.0) { 1 }  
}
}
;
}}

\newcommand{\forestVJ}{
\tikz[planar forest] {

\node [b] at (0.0, 0.0) {  } 
child {node [b] at (-0.5, 1.0) {  }  
child {node [lc] at (0.0, 1.0) { 1 }  
}
}
child {node [lc] at (0.5, 1.0) { 1 }  
}
;
}}

\newcommand{\forestWJ}{
\tikz[planar forest] {

\node [b] at (0.0, 0.0) {  } 
child {node [b] at (-1.0, 1.0) {  }  
}
child {node [lc] at (0.0, 1.0) { 1 }  
}
child {node [lc] at (1.0, 1.0) { 1 }  
}
;
}}

\newcommand{\forestXJ}{
\tikz[planar forest] {

\node [b] at (0.0, 0.0) {  } 
child {node [lc] at (-1.5, 1.0) { 1 }  
}
child {node [lc] at (-0.5, 1.0) { 1 }  
}
child {node [lc] at (0.5, 1.0) { 2 }  
}
child {node [lc] at (1.5, 1.0) { 2 }  
}
;
}}

\newcommand{\forestYJ}{
\tikz[planar forest] {

\draw (0.0,1.0) arc (90:270:0.5);
\draw (0.0,1.0) arc (90:-90:0.5);
\draw (0.0,0.0) edge (0.0,0.0);
\node [draw=none] at (0.0, 0.0) { } 
child {node [b] at (0.0, 1.0) {  } edge from parent[draw=none] 
child {node [b] at (0.0, 1.0) {  }  
child {node [lc] at (0.0, 1.0) { 1 }  
}
}
}
;
\node [lc] at (1.1, 0.0) { 1 } 
;
}}

\newcommand{\forestAK}{
\tikz[planar forest] {

\draw (0.0,1.0) arc (90:270:0.5);
\draw (0.0,1.0) arc (90:-90:0.5);
\draw (0.0,0.0) edge (0.0,0.0);
\node [draw=none] at (0.0, 0.0) { } 
child {node [b] at (0.0, 1.0) {  } edge from parent[draw=none] 
child {node [b] at (-0.5, 1.0) {  }  
}
child {node [lc] at (0.5, 1.0) { 1 }  
}
}
;
\node [lc] at (1.1, 0.0) { 1 } 
;
}}

\newcommand{\forestBK}{
\tikz[planar forest] {

\draw (0.0,1.0) arc (90:270:0.5);
\draw (0.0,1.0) arc (90:-90:0.5);
\draw (0.0,0.0) edge (0.0,0.0);
\node [draw=none] at (0.0, 0.0) { } 
child {node [b] at (0.0, 1.0) {  } edge from parent[draw=none] 
child {node [lc] at (0.0, 1.0) { 1 }  
}
}
;
\node [b] at (1.1, 0.0) {  } 
child {node [lc] at (0.0, 1.0) { 1 }  
}
;
}}

\newcommand{\forestCK}{
\tikz[planar forest] {

\draw (-0.5,1.0) edge (0.5,1.0);
\draw (-0.5,1.0) arc (90:270:0.5);
\draw (0.5,1.0) arc (90:-90:0.5);
\draw (-0.5,0.0) edge (0.5,0.0);
\node [draw=none] at (0.0, 0.0) { } 
child {node [b] at (-0.5, 1.0) {  } edge from parent[draw=none] 
}
child {node [b] at (0.5, 1.0) {  } edge from parent[draw=none] 
child {node [lc] at (0.0, 1.0) { 1 }  
}
}
;
\node [lc] at (1.6, 0.0) { 1 } 
;
}}

\newcommand{\forestDK}{
\tikz[planar forest] {

\draw (0.0,1.0) arc (90:270:0.5);
\draw (0.0,1.0) arc (90:-90:0.5);
\draw (0.0,0.0) edge (0.0,0.0);
\node [draw=none] at (0.0, 0.0) { } 
child {node [b] at (0.0, 1.0) {  } edge from parent[draw=none] 
child {node [lc] at (-1.0, 1.0) { 1 }  
}
child {node [lc] at (0.0, 1.0) { 2 }  
}
child {node [lc] at (1.0, 1.0) { 2 }  
}
}
;
\node [lc] at (1.1, 0.0) { 1 } 
;
}}

\newcommand{\forestEK}{
\tikz[planar forest] {

\draw (0.0,-0.1) edge (1.0,-0.1);
\draw (0.0,0.1) edge (1.0,0.1);
;
\node [b] at (0.0, 0.0) {  } 
;
\node [b] at (1.0, 0.0) {  } 
child {node [b] at (0.0, 1.0) {  }  
child {node [lc] at (0.0, 1.0) { 1 }  
}
}
;
\node [lc] at (2.0, 0.0) { 1 } 
;
}}

\newcommand{\forestFK}{
\tikz[planar forest] {

\node [b] at (0.0, 0.0) {  } 
child {node [b] at (0.0, 1.0) {  }  
child {node [lc] at (0.0, 1.0) { 1 }  
}
}
;
\node [lc] at (1.0, 0.0) { 1 } 
;
}}

\newcommand{\forestGK}{
\tikz[planar forest] {

\node [b] at (0.0, 0.0) {  } 
child {node [b] at (-0.5, 1.0) {  }  
}
child {node [lc] at (0.5, 1.0) { 1 }  
}
;
\node [lc] at (1.0, 0.0) { 1 } 
;
}}

\newcommand{\forestHK}{
\tikz[planar forest] {

\node [b] at (0.0, 0.0) {  } 
child {node [lc] at (0.0, 1.0) { 1 }  
}
;
\node [b] at (1.0, 0.0) {  } 
child {node [lc] at (0.0, 1.0) { 1 }  
}
;
}}

\newcommand{\forestIK}{
\tikz[planar forest] {

\node [b] at (0.0, 0.0) {  } 
child {node [lc] at (-1.0, 1.0) { 1 }  
}
child {node [lc] at (0.0, 1.0) { 2 }  
}
child {node [lc] at (1.0, 1.0) { 2 }  
}
;
\node [lc] at (1.0, 0.0) { 1 } 
;
}}

\newcommand{\forestJK}{
\tikz[planar forest] {

\draw (0.0,-0.1) edge (1.0,-0.1);
\draw (0.0,0.1) edge (1.0,0.1);
;
\node [b] at (0.0, 0.0) {  } 
;
\node [b] at (1.0, 0.0) {  } 
child {node [b] at (-0.5, 1.0) {  }  
}
child {node [lc] at (0.5, 1.0) { 1 }  
}
;
\node [lc] at (2.0, 0.0) { 1 } 
;
}}

\newcommand{\forestKK}{
\tikz[planar forest] {

\draw (0.0,-0.1) edge (1.0,-0.1);
\draw (0.0,0.1) edge (1.0,0.1);
;
\node [b] at (0.0, 0.0) {  } 
child {node [lc] at (0.0, 1.0) { 1 }  
}
;
\node [b] at (1.0, 0.0) {  } 
child {node [b] at (0.0, 1.0) {  }  
}
;
\node [lc] at (2.0, 0.0) { 1 } 
;
}}

\newcommand{\forestLK}{
\tikz[planar forest] {

\draw (0.0,-0.1) edge (1.0,-0.1);
\draw (0.0,0.1) edge (1.0,0.1);
;
\node [b] at (0.0, 0.0) {  } 
;
\node [b] at (1.0, 0.0) {  } 
child {node [lc] at (-1.0, 1.0) { 1 }  
}
child {node [lc] at (0.0, 1.0) { 1 }  
}
child {node [lc] at (1.0, 1.0) { 2 }  
}
;
\node [lc] at (2.0, 0.0) { 2 } 
;
}}

\newcommand{\forestMK}{
\tikz[planar forest] {

\draw (0.0,-0.1) edge (1.5,-0.1);
\draw (0.0,0.1) edge (1.5,0.1);
;
\node [b] at (0.0, 0.0) {  } 
child {node [lc] at (0.0, 1.0) { 2 }  
}
;
\node [b] at (1.5, 0.0) {  } 
child {node [lc] at (-0.5, 1.0) { 1 }  
}
child {node [lc] at (0.5, 1.0) { 1 }  
}
;
\node [lc] at (2.5, 0.0) { 2 } 
;
}}

\newcommand{\forestNK}{
\tikz[planar forest] {

\draw (0.0,-0.1) edge (1.5,-0.1);
\draw (0.0,0.1) edge (1.5,0.1);
;
\node [b] at (0.0, 0.0) {  } 
child {node [lc] at (0.0, 1.0) { 1 }  
}
;
\node [b] at (1.5, 0.0) {  } 
child {node [lc] at (-0.5, 1.0) { 1 }  
}
child {node [lc] at (0.5, 1.0) { 2 }  
}
;
\node [lc] at (2.5, 0.0) { 2 } 
;
}}

\newcommand{\forestOK}{
\tikz[planar forest] {

\draw (0.0,1.0) arc (90:270:0.5);
\draw (0.0,1.0) arc (90:-90:0.5);
\draw (0.0,0.0) edge (0.0,0.0);
\node [draw=none] at (0.0, 0.0) { } 
child {node [b] at (0.0, 1.0) {  } edge from parent[draw=none] 
child {node [lc] at (-0.5, 1.0) { 1 }  
}
child {node [lc] at (0.5, 1.0) { 2 }  
}
}
;
\node [lc] at (1.1, 0.0) { 1 } 
;
\node [lc] at (2.1, 0.0) { 2 } 
;
}}

\newcommand{\forestPK}{
\tikz[planar forest] {

\node [b] at (0.0, 0.0) {  } 
child {node [lc] at (-0.5, 1.0) { 1 }  
}
child {node [lc] at (0.5, 1.0) { 2 }  
}
;
\node [lc] at (1.0, 0.0) { 1 } 
;
\node [lc] at (2.0, 0.0) { 2 } 
;
}}

\newcommand{\forestQK}{
\tikz[planar forest] {

\draw (0.0,-0.1) edge (1.0,-0.1);
\draw (0.0,0.1) edge (1.0,0.1);
;
\node [b] at (0.0, 0.0) {  } 
child {node [lc] at (0.0, 1.0) { 1 }  
}
;
\node [b] at (1.0, 0.0) {  } 
child {node [lc] at (0.0, 1.0) { 2 }  
}
;
\node [lc] at (2.0, 0.0) { 1 } 
;
\node [lc] at (3.0, 0.0) { 2 } 
;
}}

\newcommand{\forestRK}{
\tikz[planar forest] {

\draw (0.0,-0.1) edge (1.0,-0.1);
\draw (0.0,0.1) edge (1.0,0.1);
;
\node [b] at (0.0, 0.0) {  } 
;
\node [b] at (1.0, 0.0) {  } 
child {node [lc] at (-0.5, 1.0) { 1 }  
}
child {node [lc] at (0.5, 1.0) { 2 }  
}
;
\node [lc] at (2.0, 0.0) { 1 } 
;
\node [lc] at (3.0, 0.0) { 2 } 
;
}}

\newcommand{\forestSK}{
\tikz[planar forest] {

\node [lc] at (0.0, 0.0) { 1 } 
;
}}

\newcommand{\forestTK}{
\tikz[planar forest] {

\node [lc] at (0.0, 0.0) { 2 } 
child {node [lc] at (-0.5, 1.0) { 3 }  
}
child {node [lc] at (0.5, 1.0) { 4 }  
}
;
}}

\newcommand{\forestUK}{
\tikz[planar forest] {

\node [lc] at (0.0, 0.0) { 2 } 
child {node [lc] at (-1.0, 1.0) { 1 }  
}
child {node [lc] at (0.0, 1.0) { 3 }  
}
child {node [lc] at (1.0, 1.0) { 4 }  
}
;
}}

\newcommand{\forestVK}{
\tikz[planar forest] {

\node [lc] at (0.0, 0.0) { 2 } 
child {node [lc] at (-0.5, 1.0) { 3 }  
child {node [lc] at (0.0, 1.0) { 1 }  
}
}
child {node [lc] at (0.5, 1.0) { 4 }  
}
;
}}

\newcommand{\forestWK}{
\tikz[planar forest] {

\node [lc] at (0.0, 0.0) { 2 } 
child {node [lc] at (-0.5, 1.0) { 3 }  
}
child {node [lc] at (0.5, 1.0) { 4 }  
child {node [lc] at (0.0, 1.0) { 1 }  
}
}
;
}}

\newcommand{\forestXK}{
\tikz[planar forest] {

\node [lc] at (0.0, 0.0) { 1 } 
child {node [lc] at (0.0, 1.0) { 2 }  
}
;
\node [lc] at (1.0, 0.0) { 3 } 
;
}}

\newcommand{\forestYK}{
\tikz[planar forest] {

\node [lc] at (0.0, 0.0) { 4 } 
child {node [lc] at (-0.5, 1.0) { 5 }  
}
child {node [lc] at (0.5, 1.0) { 6 }  
}
;
}}

\newcommand{\forestAL}{
\tikz[planar forest] {

\node [lc] at (0.0, 0.0) { 4 } 
child {node [lc] at (-0.5, 1.0) { 5 }  
}
child {node [lc] at (0.5, 1.0) { 6 }  
child {node [lc] at (-0.5, 1.0) { 1 }  
child {node [lc] at (0.0, 1.0) { 2 }  
}
}
child {node [lc] at (0.5, 1.0) { 3 }  
}
}
;
}}

\newcommand{\forestBL}{
\tikz[planar forest] {

\node [lc] at (0.0, 0.0) { 4 } 
child {node [lc] at (-0.5, 1.0) { 5 }  
child {node [lc] at (0.0, 1.0) { 1 }  
child {node [lc] at (0.0, 1.0) { 2 }  
}
}
}
child {node [lc] at (0.5, 1.0) { 6 }  
child {node [lc] at (0.0, 1.0) { 3 }  
}
}
;
}}

\newcommand{\forestCL}{
\tikz[planar forest] {

\node [lc] at (0.0, 0.0) { 4 } 
child {node [lc] at (-0.5, 1.0) { 5 }  
child {node [lc] at (0.0, 1.0) { 3 }  
}
}
child {node [lc] at (0.5, 1.0) { 6 }  
child {node [lc] at (0.0, 1.0) { 1 }  
child {node [lc] at (0.0, 1.0) { 2 }  
}
}
}
;
}}

\newcommand{\forestDL}{
\tikz[planar forest] {

\node [lc] at (0.0, 0.0) { 4 } 
child {node [lc] at (-0.5, 1.0) { 5 }  
child {node [lc] at (-0.5, 1.0) { 1 }  
child {node [lc] at (0.0, 1.0) { 2 }  
}
}
child {node [lc] at (0.5, 1.0) { 3 }  
}
}
child {node [lc] at (0.5, 1.0) { 6 }  
}
;
}}

\newcommand{\forestEL}{
\tikz[planar forest] {

\node [lc] at (0.0, 0.0) { 4 } 
child {node [lc] at (-1.0, 1.0) { 1 }  
child {node [lc] at (0.0, 1.0) { 2 }  
}
}
child {node [lc] at (0.0, 1.0) { 5 }  
}
child {node [lc] at (1.0, 1.0) { 6 }  
child {node [lc] at (0.0, 1.0) { 3 }  
}
}
;
}}

\newcommand{\forestFL}{
\tikz[planar forest] {

\node [lc] at (0.0, 0.0) { 4 } 
child {node [lc] at (-1.0, 1.0) { 1 }  
child {node [lc] at (0.0, 1.0) { 2 }  
}
}
child {node [lc] at (0.0, 1.0) { 5 }  
child {node [lc] at (0.0, 1.0) { 3 }  
}
}
child {node [lc] at (1.0, 1.0) { 6 }  
}
;
}}

\newcommand{\forestGL}{
\tikz[planar forest] {

\node [lc] at (0.0, 0.0) { 4 } 
child {node [lc] at (-1.0, 1.0) { 3 }  
}
child {node [lc] at (0.0, 1.0) { 5 }  
}
child {node [lc] at (1.0, 1.0) { 6 }  
child {node [lc] at (0.0, 1.0) { 1 }  
child {node [lc] at (0.0, 1.0) { 2 }  
}
}
}
;
}}

\newcommand{\forestHL}{
\tikz[planar forest] {

\node [lc] at (0.0, 0.0) { 4 } 
child {node [lc] at (-1.0, 1.0) { 3 }  
}
child {node [lc] at (0.0, 1.0) { 5 }  
child {node [lc] at (0.0, 1.0) { 1 }  
child {node [lc] at (0.0, 1.0) { 2 }  
}
}
}
child {node [lc] at (1.0, 1.0) { 6 }  
}
;
}}

\newcommand{\forestIL}{
\tikz[planar forest] {

\node [lc] at (0.0, 0.0) { 4 } 
child {node [lc] at (-1.5, 1.0) { 1 }  
child {node [lc] at (0.0, 1.0) { 2 }  
}
}
child {node [lc] at (-0.5, 1.0) { 3 }  
}
child {node [lc] at (0.5, 1.0) { 5 }  
}
child {node [lc] at (1.5, 1.0) { 6 }  
}
;
}}

%% file: references.bib
@article{ketcheson2023computing,
    title = {Computing with {B}-series},
    author = {Ketcheson, David I and Ranocha, Hendrik},
    journal = {ACM Transactions on Mathematical Software},
    volume = {49},
    number = {2},
    year = {2023},
    month = {06},
    doi = {10.1145/3573384},
    eprint = {2111.11680},
    eprinttype = {arXiv},
    eprintclass = {math.NA},
}

@misc{ranocha2021bseries,
    title = {{BSeries.jl}: {C}omputing with {B}-series in {J}ulia},
    author = {Ranocha, Hendrik and Ketcheson, David I},
    year = {2021},
    month = {09},
    howpublished = {\url{https://github.com/ranocha/BSeries.jl}},
    doi = {10.5281/zenodo.5534602},
}

@misc{BSeries.py,
    AUTHOR = {Ranocha, Hendrik and Ketcheson, David I},
    TITLE = {BSeries.py},
    YEAR = {2021},
    HOWPUBLISHED = {\url{https://github.com/ketch/BSeries}},
}

@mastersthesis{Sundklakk15pybs,
    AUTHOR = {Sundklakk, Henrik Sperre},
    TITLE = {A library for computing with trees and {B}-{S}eries},
    SCHOOL = {NTNU},
    YEAR = {2015},
    NOTE = {Supervisor: B. Owren},
}

@misc{pybs,
    AUTHOR = {Sundklakk, Henrik Sperre},
    TITLE = {pyBS},
    YEAR = {2015},
    HOWPUBLISHED = {\url{https://github.com/henriksu/pybs}},
}

@article{hairer2014,
    title = {A Theory of Regularity Structures},
    author = {Hairer, M.},
    year = {2014},
    month = nov,
    journal = {Inventiones mathematicae},
    volume = {198},
    number = {2},
    pages = {269--504},
    issn = {1432-1297},
    doi = {10.1007/s00222-014-0505-4},
    urldate = {2025-05-02},
    langid = {english},
}

@article{gubinelli2010,
    title = {Ramification of Rough Paths},
    author = {Gubinelli, Massimiliano},
    year = {2010},
    month = feb,
    journal = {Journal of Differential Equations},
    volume = {248},
    number = {4},
    pages = {693--721},
    issn = {0022-0396},
    doi = {10.1016/j.jde.2009.11.015},
    urldate = {2025-05-02},
}

@article{BronascoHAS24,
    title = {Hopf Algebra Structures for the Backward Error Analysis of Ergodic
             Stochastic Differential Equations},
    author = {Bronasco, Eugen and Busnot Laurent, Adrien},
    year = 2026,
    month = mar,
    journal = {Numerische Mathematik},
    issn = {0945-3245},
    doi = {10.1007/s00211-026-01533-7},
    urldate = {2026-04-21},
    langid = {english},
}

@book{HairerSOD93,
    title = {Solving {{Ordinary Differential Equations I}}},
    author = {Hairer, Ernst and N{\o}rsett, Syvert P. and Wanner, Gerhard},
    year = 1993,
    series = {Springer {{Series}} in {{Computational Mathematics}}},
    volume = {8},
    publisher = {Springer},
    address = {Berlin, Heidelberg},
    doi = {10.1007/978-3-540-78862-1},
    urldate = {2025-11-12},
    copyright = {http://www.springer.com/tdm},
    isbn = {978-3-540-56670-0 978-3-540-78862-1},
    langid = {english},
}

@article{Cayley1857,
    author = {Cayley, A.},
    title = {On the theory of the analytical forms called trees},
    journal = {Philosophical Magazine},
    volume = {13},
    pages = {172--176},
    year = {1857},
    doi = {10.1080/14786445708642866},
}

@article{chapotonPreLieAlgebrasRooted2001,
    title = {Pre-{{Lie}} Algebras and the Rooted Trees Operad},
    author = {Chapoton, Fr{\'e}d{\'e}ric and Livernet, Muriel},
    year = 2001,
    month = jan,
    journal = {International Mathematics Research Notices},
    volume = {2001},
    number = {8},
    pages = {395--408},
    issn = {1073-7928},
    doi = {10.1155/S1073792801000198},
    urldate = {2025-05-25},
}

@article{oudomLieEnvelopingAlgebra2008,
    title = {On the {{Lie}} Enveloping Algebra of a Pre-{{Lie}} Algebra},
    author = {Oudom, J.-M. and Guin, D.},
    year = 2008,
    month = aug,
    journal = {Journal of K-Theory},
    volume = {2},
    number = {1},
    pages = {147--167},
    issn = {1865-5394, 1865-2433},
    doi = {10.1017/is008001011jkt037},
    urldate = {2025-05-25},
    langid = {english},
}

@article{ChartierASB10,
    title = {Algebraic {{Structures}} of {{B-series}}},
    author = {Chartier, Philippe and Hairer, Ernst and Vilmart, Gilles},
    year = 2010,
    month = aug,
    journal = {Foundations of Computational Mathematics},
    volume = {10},
    number = {4},
    pages = {407--427},
    issn = {1615-3383},
    doi = {10.1007/s10208-010-9065-1},
    urldate = {2025-12-26},
    langid = {english},
}

@article{calaqueTwoInteractingHopf2011,
    title = {Two Interacting {{Hopf}} Algebras of Trees},
    author = {Calaque, Damien and {Ebrahimi-Fard}, Kurusch and Manchon,
              Dominique},
    year = 2011,
    month = aug,
    journal = {Advances in Applied Mathematics},
    volume = {47},
    number = {2},
    eprint = {0806.2238},
    primaryclass = {math},
    pages = {282--308},
    issn = {01968858},
    doi = {10.1016/j.aam.2009.08.003},
    urldate = {2025-04-24},
    archiveprefix = {arXiv},
}

@article{laurentOrderConditionsSampling2022,
    title = {Order Conditions for Sampling the Invariant Measure of Ergodic
             Stochastic Differential Equations on Manifolds},
    author = {Laurent, Adrien and Vilmart, Gilles},
    year = 2022,
    month = jun,
    journal = {Foundations of Computational Mathematics},
    volume = {22},
    number = {3},
    pages = {649--695},
    issn = {1615-3383},
    doi = {10.1007/s10208-021-09495-y},
    urldate = {2025-05-25},
    langid = {english},
}

@article{BronascoEBS25,
    title = {Exotic {{B-Series}} and {{S-Series}}: {{Algebraic Structures}} and
             {{Order Conditions}} for {{Invariant Measure Sampling}}},
    shorttitle = {Exotic {{B-Series}} and {{S-Series}}},
    author = {Bronasco, Eugen},
    year = 2025,
    month = feb,
    journal = {Foundations of Computational Mathematics},
    volume = {25},
    number = {1},
    pages = {271--301},
    issn = {1615-3383},
    doi = {10.1007/s10208-023-09638-3},
    urldate = {2026-04-21},
    langid = {english},
}

@inproceedings{floystadUniversalPreLieRinehartAlgebras2021,
    title = {The Universal Pre-{{Lie}}–{{Rinehart}} Algebras of Aromatic Trees},
    booktitle = {Geometric and {{Harmonic Analysis}} on {{Homogeneous Spaces}}
                 and {{Applications}}},
    author = {Fløystad, Gunnar and Manchon, Dominique and Munthe-Kaas, Hans Z.},
    editor = {Baklouti, Ali and Ishi, Hideyuki},
    year = {2021},
    pages = {137--159},
    publisher = {Springer International Publishing},
    location = {Cham},
    doi = {10.1007/978-3-030-78346-4_9},
    isbn = {978-3-030-78346-4},
    langid = {english},
}

@article{DotsenkoGBO10,
    title = {Gröbner Bases for Operads},
    author = {Dotsenko, Vladimir and Khoroshkin, Anton},
    year = {2010},
    journal = {Duke Mathematical Journal},
    volume = {153},
    number = {2},
    pages = {363--396},
    publisher = {Duke University Press},
    issn = {0012-7094, 1547-7398},
    doi = {10.1215/00127094-2010-026},
    url = {
           https://projecteuclid.org/journals/duke-mathematical-journal/volume-153/issue-2/Gr%c3%b6bner-bases-for-operads/10.1215/00127094-2010-026.full
           },
    urldate = {2026-01-05},
}

@article{Lyons1998,
    author = {Terry J. Lyons},
    title = {Differential equations driven by rough signals},
    journal = {Revista Matem{\'a}tica Iberoamericana},
    volume = {14},
    number = {2},
    pages = {215--310},
    year = {1998},
}

@article{BrunedARR19,
    title = {Algebraic Renormalisation of Regularity Structures},
    author = {Bruned, Y. and Hairer, M. and Zambotti, L.},
    year = 2019,
    month = mar,
    journal = {Inventiones mathematicae},
    volume = {215},
    number = {3},
    pages = {1039--1156},
    issn = {1432-1297},
    doi = {10.1007/s00222-018-0841-x},
    urldate = {2026-03-24},
    langid = {english},
}

@article{butcherCoefficientsStudyRungeKutta1963,
    title = {Coefficients for the Study of {{Runge-Kutta}} Integration Processes
             },
    author = {Butcher, J. C.},
    year = 1963,
    month = may,
    journal = {Journal of the Australian Mathematical Society},
    volume = {3},
    number = {2},
    pages = {185--201},
    issn = {0004-9735},
    doi = {10.1017/S1446788700027932},
    urldate = {2025-05-25},
    langid = {english},
}

@article{MuruaHAR06,
    title = {The {{Hopf Algebra}} of {{Rooted Trees}}, {{Free Lie Algebras}},
             and {{Lie Series}}},
    author = {Murua, A.},
    year = 2006,
    month = nov,
    journal = {Foundations of Computational Mathematics},
    volume = {6},
    number = {4},
    pages = {387--426},
    issn = {1615-3383},
    doi = {10.1007/s10208-003-0111-0},
    urldate = {2026-02-12},
    langid = {english},
}

@book{LodayAO12,
    title = {Algebraic {{Operads}}},
    author = {Loday, Jean-Louis and Vallette, Bruno},
    year = 2012,
    series = {Grundlehren Der Mathematischen {{Wissenschaften}}},
    volume = {346},
    publisher = {Springer},
    address = {Berlin, Heidelberg},
    doi = {10.1007/978-3-642-30362-3},
    urldate = {2026-03-24},
    copyright = {
                 https://www.springernature.com/gp/researchers/text-and-data-mining
                 },
    isbn = {978-3-642-30361-6 978-3-642-30362-3},
    langid = {english},
}

@article{bogfjellmoHamiltonianBseriesLie2017,
    title = {Hamiltonian {{B-series}} and a {{Lie}} Algebra of Non-Rooted Trees},
    author = {Bogfjellmo, Geir and Curry, Charles and Manchon, Dominique},
    year = 2017,
    month = jan,
    journal = {Numerische Mathematik},
    volume = {135},
    number = {1},
    pages = {97--112},
    issn = {0945-3245},
    doi = {10.1007/s00211-016-0792-3},
    urldate = {2025-05-27},
    langid = {english},
}

@article{Munthe-KaasLTR95,
    title = {Lie-{{Butcher}} Theory for {{Runge-Kutta}} Methods},
    author = {{Munthe-Kaas}, Hans},
    year = 1995,
    month = dec,
    journal = {BIT Numerical Mathematics},
    volume = {35},
    number = {4},
    pages = {572--587},
    issn = {1572-9125},
    doi = {10.1007/BF01739828},
    urldate = {2025-07-23},
    langid = {english},
}

@misc{munthe-kaasHopfAlgebraicStructure2006,
    title = {On the {{Hopf Algebraic Structure}} of {{Lie Group Integrators}}},
    author = {{Munthe-Kaas}, H. Z. and Wright, W. M.},
    year = 2006,
    month = mar,
    number = {arXiv:math/0603023},
    eprint = {math/0603023},
    publisher = {arXiv},
    doi = {10.48550/arXiv.math/0603023},
    urldate = {2025-05-23},
    archiveprefix = {arXiv},
}

@article{BurrageBurrage,
    title = {Order {{Conditions}} of {{Stochastic Runge--Kutta Methods}} by {{
             B-Series}}},
    author = {Burrage, K. and Burrage, P. M.},
    year = 2000,
    journal = {SIAM Journal on Numerical Analysis},
    volume = {38},
    number = {5},
    pages = {1626--1646},
    urldate = {2026-03-24},
    langid = {english},
}

@article{RosslerRTA06,
    title = {Rooted {{Tree Analysis}} for {{Order Conditions}} of {{Stochastic
             Runge-Kutta Methods}} for the {{Weak Approximation}} of {{Stochastic
             Differential Equations}}},
    author = {R{\"o}{\ss}ler, Andreas},
    year = 2006,
    month = mar,
    journal = {Stochastic Analysis and Applications},
    volume = {24},
    number = {1},
    pages = {97--134},
    publisher = {Taylor \& Francis},
    issn = {0736-2994},
    doi = {10.1080/07362990500397699},
    urldate = {2026-03-24},
}

@article{LaurentEAB20,
    title = {Exotic Aromatic {{B-series}} for the Study of Long Time Integrators
             for a Class of Ergodic {{SDE}}s},
    author = {Laurent, Adrien and Vilmart, Gilles},
    year = 2020,
    month = jan,
    journal = {Mathematics of Computation},
    volume = {89},
    number = {321},
    pages = {169--202},
    issn = {0025-5718, 1088-6842},
    doi = {10.1090/mcom/3455},
    urldate = {2025-09-09},
    langid = {english},
}

@article{iserlesBSeriesMethodsCannot2007,
    title = {B-{{Series}} Methods Cannot Be Volume-Preserving},
    author = {Iserles, A. and Quispel, G.R.W. and Tse, P.S.P.},
    year = 2007,
    month = jun,
    journal = {BIT Numerical Mathematics},
    volume = {47},
    number = {2},
    pages = {351--378},
    issn = {1572-9125},
    doi = {10.1007/s10543-006-0114-8},
    urldate = {2025-05-25},
    langid = {english},
}

@article{ChartierPFI07,
    title = {Preserving First Integrals and Volume Forms of Additively Split
             Systems},
    author = {Chartier, Philippe and Murua, Ander},
    year = 2007,
    month = apr,
    journal = {IMA Journal of Numerical Analysis},
    volume = {27},
    number = {2},
    pages = {381--405},
    issn = {0272-4979},
    doi = {10.1093/imanum/drl039},
    urldate = {2026-03-24},
}

@article{munthe-kaasAromaticButcherSeries2016,
    title = {Aromatic {{Butcher Series}}},
    author = {{Munthe-Kaas}, Hans and Verdier, Olivier},
    year = 2016,
    month = feb,
    journal = {Foundations of Computational Mathematics},
    volume = {16},
    number = {1},
    pages = {183--215},
    issn = {1615-3383},
    doi = {10.1007/s10208-015-9245-0},
    urldate = {2025-05-25},
    langid = {english},
}

@inproceedings{MokhovAGC17,
    title = {Algebraic Graphs with Class (Functional Pearl)},
    booktitle = {Proceedings of the 10th {{ACM SIGPLAN International Symposium}}
                 on {{Haskell}}},
    author = {Mokhov, Andrey},
    year = 2017,
    month = sep,
    series = {Haskell 2017},
    pages = {2--13},
    publisher = {Association for Computing Machinery},
    address = {New York, NY, USA},
    doi = {10.1145/3122955.3122956},
    urldate = {2026-03-24},
    isbn = {978-1-4503-5182-9},
}

@article{bronascoEfficientLangevinSampling2025,
    title = {Efficient {{Langevin}} Sampling with Position-Dependent Diffusion},
    author = {Bronasco, Eugen and Leimkuhler, Benedict and Phillips, Dominic and
              Vilmart, Gilles},
    year = 2025,
    month = jan,
    journal = {submitted for publication},
    number = {arXiv:2501.02943},
    eprint = {2501.02943},
    primaryclass = {math},
    publisher = {arXiv},
    doi = {10.48550/arXiv.2501.02943},
    urldate = {2025-04-26},
    archiveprefix = {arXiv},
    copyright = {All rights reserved},
}

@book{ButcherBAA21,
    title = {B-{{Series}}: {{Algebraic Analysis}} of {{Numerical Methods}}},
    shorttitle = {B-{{Series}}},
    author = {Butcher, John C.},
    year = 2021,
    series = {Springer {{Series}} in {{Computational Mathematics}}},
    volume = {55},
    publisher = {Springer International Publishing},
    address = {Cham},
    doi = {10.1007/978-3-030-70956-3},
    urldate = {2026-04-21},
    copyright = {http://www.springer.com/tdm},
    isbn = {978-3-030-70955-6 978-3-030-70956-3},
    langid = {english},
}

@phdthesis{BronascoThesis,
    TITLE = {Algebraic structures and numerical methods for invariant measure
             sampling of {L}angevin dynamics},
    AUTHOR = {Bronasco, Eugen},
    SCHOOL = {University of Geneva},
    YEAR = {2025},
    TYPE = {Thesis},
    DOI = {10.13097/archive-ouverte/unige:185162},
}

@misc{bronasco_zenodo,
    AUTHOR = {Bronasco, Eugen},
    TITLE = {arboretum.hs},
    YEAR = 2026,
    publisher = {Zenodo},
    VERSION = {0.2.0.0},
    DOI = {10.5281/zenodo.19737034},
    URL = {https://doi.org/10.5281/zenodo.19737034},
    HOWPUBLISHED = {\url{https://doi.org/10.5281/zenodo.19737034}},
}

@article{Butcher_72,
    AUTHOR = {Butcher, J. C.},
    TITLE = {An algebraic theory of integration methods},
    JOURNAL = {Math. Comp.},
    FJOURNAL = {Mathematics of Computation},
    VOLUME = {26},
    YEAR = {1972},
    PAGES = {79--106},
    ISSN = {0025-5718,1088-6842},
    MRCLASS = {65L99},
    MRNUMBER = {305608},
    MRREVIEWER = {D.\ B.\ Hunter},
    DOI = {10.2307/2004720},
    URL = {https://doi.org/10.2307/2004720},
}

@article{connesHopfAlgebrasRenormalization1998,
    title = {Hopf {{Algebras}}, {{Renormalization}} and {{Noncommutative
             Geometry}}},
    author = {Connes, Alain and Kreimer, Dirk},
    year = 1998,
    month = dec,
    journal = {Communications in Mathematical Physics},
    volume = {199},
    number = {1},
    pages = {203--242},
    issn = {1432-0916},
    doi = {10.1007/s002200050499},
    urldate = {2025-05-25},
    langid = {english},
}

@article{BrouderRKM00,
    title = {Runge--{{Kutta}} Methods and Renormalization},
    author = {Brouder, {\relax Ch}.},
    year = 2000,
    month = feb,
    journal = {The European Physical Journal C - Particles and Fields},
    volume = {12},
    number = {3},
    pages = {521--534},
    issn = {1434-6052},
    doi = {10.1007/s100529900235},
    urldate = {2026-04-28},
    langid = {english},
}


%% file: zotero.bib
@article{bogfjellmo2019,
    title = {Algebraic Structure of Aromatic {{B-series}}},
    author = {Bogfjellmo, Geir},
    year = {2019},
    journal = {Journal of Computational Dynamics},
    volume = {6},
    number = {2},
    pages = {199--222},
    publisher = {Journal of Computational Dynamics},
    issn = {2158-2491},
    doi = {10.3934/jcd.2019010},
    urldate = {2025-05-25},
    copyright = {http://creativecommons.org/licenses/by/3.0/},
    langid = {english},
}

@article{hairer1974,
    title = {On the {{Butcher}} Group and General Multi-Value Methods},
    author = {Hairer, E. and Wanner, G.},
    year = {1974},
    month = mar,
    journal = {Computing},
    volume = {13},
    number = {1},
    pages = {1--15},
    issn = {1436-5057},
    doi = {10.1007/BF02268387},
    urldate = {2025-05-25},
    langid = {english},
}

@book{HairerWannerGNI,
    title = {Geometric {{Numerical Integration}}: {{Structure-Preserving
             Algorithms}} for {{Ordinary Differential Equations}}},
    shorttitle = {Geometric {{Numerical Integration}}},
    author = {Hairer, Ernst and Lubich, Christian and Wanner, Gerhard},
    year = {2010},
    month = mar,
    publisher = {Springer Berlin Heidelberg},
    googlebooks = {ssrFQQAACAAJ},
    isbn = {978-3-642-05157-9},
    langid = {english},
}

@inproceedings{munthe-kaas2018,
    title = {Lie--{{Butcher Series}}, {{Geometry}}, {{Algebra}} and {{
             Computation}}},
    booktitle = {Discrete {{Mechanics}}, {{Geometric Integration}} and {{Lie}}--
                 {{Butcher Series}}},
    author = {{Munthe-Kaas}, Hans Z. and F{\o}llesdal, Kristoffer K.},
    editor = {{Ebrahimi-Fard}, Kurusch and Barbero Li{\~n}{\'a}n, Mar{\'i}a},
    year = {2018},
    pages = {71--113},
    publisher = {Springer International Publishing},
    address = {Cham},
    doi = {10.1007/978-3-030-01397-4_3},
    isbn = {978-3-030-01397-4},
    langid = {english},
}
